%% file: main.tex
\documentclass[sigconf]{acmart}
\usepackage{subcaption} 

\setcopyright{acmlicensed}
\copyrightyear{2018}
\acmYear{2018}
\acmDOI{XXXXXXX.XXXXXXX}

\acmConference[Conference acronym 'XX]{Make sure to enter the correct
  conference title from your rights confirmation emai}{June 03--05,
  2018}{Woodstock, NY}

\acmISBN{978-1-4503-XXXX-X/18/06}

\usepackage{graphicx} 
\usepackage{multirow}

\newenvironment {squishlist}
{\begin{list}{$\bullet$}
  { \setlength{\itemsep}{0pt}
     \setlength{\parsep}{3pt}
     \setlength{\topsep}{3pt}
     \setlength{\partopsep}{0pt}
     \setlength{\leftmargin}{1.5em}
     \setlength{\labelwidth}{1em}
     \setlength{\labelsep}{0.5em} } }
{\end{list}}

\setcopyright{none}
\begin{document}

\title{Analyzing Patterns and Influence of Advertising in Print Newspapers}


\author{N Harsha Vardhan}
\email{nemani.v@research.iiit.ac.in}
\affiliation{
  \institution{International Institute of Information Technology}
  \city{Hyderabad}
  \country{India}
}

\author{Ponnurangam Kumaraguru}
\email{pk.guru@iiit.ac.in}
\affiliation{
  \institution{International Institute of Information Technology}
  \city{Hyderabad}
  \country{India}
}

\author{Kiran Garimella}
\email{kg766@rutgers.edu}
\affiliation{
  \institution{Rutgers University}
  \city{New Brunswick}
  \country{USA}
}

\renewcommand{\shortauthors}{Vardhan et al.}

\begin{abstract}
This paper investigates advertising practices in print newspapers across India using a novel data-driven approach. We develop a pipeline employing image processing and OCR techniques to extract articles and advertisements from digital versions of print newspapers with high accuracy. Applying this methodology to five popular newspapers that span multiple regions and three languages—English, Hindi, and Telugu—we assembled a dataset of more than 12,000 editions containing several hundred thousand advertisements. Collectively, these newspapers reach a readership of over 100 million people. Using this extensive dataset, we conduct a comprehensive analysis to answer key questions about print advertising: who advertises, what they advertise, when they advertise, where they place their ads, and how they advertise. Our findings reveal significant patterns, including the consistent level of print advertising over the past six years despite declining print circulation, the overrepresentation of company ads on prominent pages, and the disproportionate revenue contributed by government ads.
Furthermore, we examine whether advertising in a newspaper influences the coverage an advertiser receives. Through regression analyses on coverage volume and sentiment, we find strong evidence supporting this hypothesis for corporate advertisers. The results indicate a clear trend where increased advertising correlates with more favorable and extensive media coverage, a relationship that remains robust over time and across different levels of advertiser popularity.

\end{abstract}




\maketitle

\section{Introduction}

In this paper, we investigate advertising in \textit{print} newspapers across India. Despite the exponential growth of digital news consumption, print newspapers continue to hold a significant position in the media landscape, both globally and within India.
Internationally, print media remains a vital source of information for many. According to a study by the Pew Research Center, 53\% of Americans who prefer newspapers for local news opt for the print version over digital formats \cite{PewLocal2023}. The enduring trust in print media further emphasizes its importance. The Reuters Institute Digital News Report indicates that 62\% of respondents trust print newspapers, compared to 44\% for news websites \cite{ReutersTrust2023}. This heightened level of trust is often attributed to the perceived editorial rigor and accountability inherent in print journalism.
Economically, print media continues to be a substantial revenue generator for major news outlets. As of 2023, print circulation and advertising contributed approximately 60-70\% of total revenue for several leading U.S. newspapers \cite{NewsRevenue2023}. The New York Times, for example, reported that about 30\% of its total revenue was derived from print subscriptions and advertising \cite{NYTFinancial2023}. Additionally, the effectiveness of print advertising bolsters its continued relevance. A study by NewsMediaWorks found that print newspaper advertisements were 1.5 times more effective in driving purchase intent compared to digital ads \cite{NewsMediaWorks2022}.

In the Indian context, the significance of print media is even more pronounced. The Indian Readership Survey of 2020~\cite{IndianReadershipSurvey}, the primary source for newspaper readership data in India, reveals that India is the second-largest newspaper market globally, with over 110 million copies sold daily. Print media commands about 20\% of India's total advertising expenditure, a stark contrast to the global average of just 4\%~\cite{indiatimesStrongOutlook}. The industry is poised for robust growth in 2025, driven by factors such as elections, with print advertising revenue anticipated to reach an all-time high.

Previous research analyzing advertising at scale has primarily focused on digital newspapers~\cite{sridhar2015online} or has been conducted on a small scale~\cite{lindstadt2011newspaper}. This is largely due to the difficulty of obtaining large-scale data on print advertisements. Such data is either exclusively held by the newspapers themselves, making cross-paper comparisons impossible, or it is expensive to acquire from third-party sellers~\cite{pressgazetteNewspaperCirculations}, who may not even have data from countries like India.

In this paper, we obtain digital versions of print newspapers and use advances in image processing and Optical Character Recognition (OCR) techniques to develop a pipeline to extract articles and ads from print newspapers.
We obtain data from 5 popular newspapers covering multiple regions and 3 languages from India (English, Hindi, and Telugu) with an aggregated reach of over 100 million people.
We obtain articles spanning multiple years from these newspapers and apply our pipeline to extract with high confidence, \textit{all} ads published in the paper along with their content.

Using this rich dataset, we answer, for the first time, \textit{who} advertises, \textit{what} they advertise, \textit{when} (e.g. what time of the week, time of year, etc), \textit{where} (e.g. on what page) and how (e.g. display ads vs. text ads). 
We specifically focus on two important categories of advertisers -- corporates and government, which together account for roughly half of the total number of advertisements and advertising expenditure. We reveal some interesting patterns in print advertising.
For instance, despite the fact that print circulation is dropping significantly~\cite{PewDemographics2023}, print advertising remained consistent over the past 6 years. We uncover key patterns of advertisements by various companies and governments, including the over representation of company ads on prominent pages and the disproportionate revenue that government ads contribute to print advertising.

Moreover, using this data, we try to answer an important research question at scale -- does advertising in a newspaper provide favorable coverage to the advertiser?
We answer this question by regressing both the coverage volume and sentiment for various entities (companies and government). We find strong evidence for this hypothesis in the case of companies, indicating a clear trend in how advertising influences coverage as well as sentiment. We show that this relationship is robust to time and popularity.

The findings help us understand and analyze print media at scale for the first time providing a key source to complement analysis of digital media.

\noindent\textbf{Contributions}.
In this paper, we make several key contributions to the study of print media advertising and its influence on news coverage:

\begin{squishlist}

\item We introduce a novel pipeline capable of extracting and analyzing all advertisements from print newspapers. This pipeline built on advances in state of the art OCR and image segmentation and is designed to handle large volumes of data efficiently, enabling extensive analysis of advertising content at an unprecedented scale.

\item Applying our pipeline, we compiled a unique and extensive dataset encompassing five different newspapers in India. This dataset covers four regions, three languages, and includes 12,358 editions, resulting in hundreds of thousands of articles and advertisements. The diversity of the dataset offers a comprehensive representation of print media advertising across different linguistic and regional contexts.

\item We provide a thorough examination of advertising strategies by addressing the fundamental questions of who advertises, what is advertised, how, where, and when. This multifaceted analysis sheds light on the behaviors and preferences of both government and corporate advertisers, revealing patterns in ad placement, timing, content, and size.

\item We study the effect of advertising on media coverage by analyzing both the tone and volume of articles related to advertisers. Our findings demonstrate a clear relationship between advertising expenditure and the nature of coverage received, particularly for corporate entities. This contributes to the understanding of how advertising revenue may influence editorial content and potentially lead to media bias.

\item Finally, we provide the code and datasets used in our approach to the community.\footnote{\url{https://github.com/harsha20032020/Code_and_Data_for_Analyzing_Patterns_and_Influence_of_Advertising_in_Print_Newspaper}} The code can help extend our analysis beyond the Indian context and the dataset can help further analysis of print news papers and their coverage.

\end{squishlist}


In an era of increasing concerns about media bias, our dataset and analysis offer critical insights into the interplay between advertisers and media outlets. By enhancing transparency and understanding of advertising practices and their potential influence on news content, we contribute to the broader discourse on media integrity, informed citizenship, and the health of democratic processes.
The methodologies and tools developed in our paper are not limited to the context of (Indian) newspapers but can be adapted to processing structured image data, thus facilitating further research into the dynamics between advertising and media coverage globally.




\section{Background and Related Work}

\noindent\textbf{Media Bias and Advertising Influence}.
The media is often regarded as the ``Fourth Estate,'' underscoring its vital role in upholding democratic societies by providing unbiased and factual reporting. This concept emphasizes the media's responsibility to act as a watchdog, holding power structures accountable and fostering informed public discourse. According to McChesney~\cite{mcchesney2004problem}, a healthy democracy necessitates a free and independent press that delivers diverse viewpoints without undue influence from external forces. However, maintaining editorial independence has become increasingly challenging due to commercial pressures \cite{siles} and the consolidation of media ownership.
Bagdikian~\cite{bagdikian2004h} highlights how media conglomerates can compromise journalistic integrity by prioritizing profit over the public interest, leading to homogenized content and potential biases. The tension between upholding democratic values and meeting commercial objectives creates a dilemma for media outlets. Hamilton~\cite{hamilton2004all} explores how economic incentives influence news production, suggesting that market demands often shape what gets reported rather than purely journalistic considerations. In this context, auditing newspapers is crucial to ensure they adhere to ethical standards and continue to serve their role as the Fourth Estate.

The imperative for profitability significantly influences editorial decisions within media organizations. \citet{picard2011economics} examines the economics of media companies, highlighting that revenue generation—particularly through advertising—often takes precedence over journalistic ideals. This commercial focus can lead to content that appeals to advertisers or attracts larger audiences, potentially at the expense of investigative journalism or critical reporting.
\citet{croteau2006business} discuss how media outlets navigate the balance between maintaining editorial integrity and satisfying corporate interests. They argue that dependence on advertising revenue can result in self-censorship or the avoidance of topics that might alienate advertisers. This balancing act affects the type of news that gets published, as media companies strive to attract and retain advertisers while attempting to uphold the standards of independent journalism. Understanding these dynamics is essential for analyzing how economic factors shape media content and the implications for democratic discourse.

\begin{table*}
\centering
\caption{Newspaper Data Comparison: The columns include the newspaper source, the time period of data collection, and the cities or zones covered. Here, ``\#'' denotes ``number of.'' The number of editions represents the total editions (each daily issue of a newspaper specific to a city and source is considered an edition) analyzed, while the number of pages is the sum of pages in each edition.}
\label{tab:newspaper_comparison}
\resizebox{\textwidth}{!}{%
\begin{tabular}{l|l|l|c|c|c|c}
\hline
\textbf{Source}          & \textbf{Time Period}           & \textbf{Cities / Zones}                         & \textbf{\# Editions}     & \textbf{\# Pages}     & \textbf{\# Articles}   & \textbf{\# Ads}        \\ \hline
Hindustan Times & July 2019 - June 2024 & Delhi, Mumbai                   & 3,547                  & 71,130              & 442,300              & 150,898              \\ \hline
Times Of India  & Dec 2021 - Jun 2024   & Delhi, Mumbai, Chennai, Kolkata & 3,724                  & 123,638             & 800,171              & 465,435              \\ \hline
Telegraph       & May 2018 - Feb 2024   & Kolkata                         & 2,077                  & 29,556              & 159,081              & 93,928               \\ \hline
Dainik Bhasker  &      Sept 2021 - Jun 2024               &       Delhi                          &    1,016                    &    13,995                 &    56,031                  &        14,862              \\ \hline
Sakshi          &      Oct 2022 - Jun 2024              &         Hyderabad, Telangana, Andhra Pradesh                        &       1,994                 &        32,465             &         162,232             &   111,250                   \\ \hline
\end{tabular}%
}
\end{table*}

\noindent\textbf{Media Bias Favoring Advertisers}.
 Empirical studies have provided evidence that advertising revenue can influence editorial content, resulting in media bias favoring advertisers. \citet{reuters} examined the financial media and found that publications tend to offer more favorable coverage to firms that advertise with them. Their analysis revealed a positive correlation between the amount of advertising purchased by a company and the tone of editorial content it receives, suggesting that editors may compromise journalistic objectivity to maintain advertising relationships.
Similarly, \citet{ditella} investigated the impact of government advertising on media coverage of corruption scandals in Argentina. They discovered that newspapers receiving substantial government advertising were less likely to report on corruption cases involving government officials. This dependence on advertising revenue from influential entities can lead to the suppression or under reporting of negative news about those entities. 
Similar trends have also been observed in India~\cite{neyazi2011india,dasgupta2017tackling}.

Ownership structures and corporate interests significantly influence media content, often leading to biased reporting. \citet{gentzkowshapiro} analyzed U.S. daily newspapers and found that the slant of news coverage is affected by both the ideological preferences of the audience and the ownership of the media outlet. Their study suggests that media owners may cater to audience biases to maximize circulation and profits, resulting in a consistent content slant that aligns with specific viewpoints and potentially limits the diversity of perspectives available to the public.

\citet{gilens2000corporate} explored how corporate ownership affects news bias by examining newspaper coverage of the 1996 Telecommunications Act. They found that newspapers owned by companies with significant television interests provided more favorable coverage of the Act, which stood to benefit their corporate holdings. This indicates that corporate interests can directly influence editorial decisions, leading to reporting that serves the owners' financial agendas. 
Government entities and large corporations often use advertising spending as a tool to influence media coverage, potentially leading to favorable reporting or the suppression of negative news. \citet{prat2013political} discussed the political economy of mass media, emphasizing how governments can manipulate news content through financial means such as advertising budgets. Their analysis highlighted that media outlets reliant on government advertising may avoid negative reporting on governmental actions to secure continued funding, thereby compromising journalistic independence.

\noindent\textbf{Challenges in print media analysis}.
Obtaining large-scale data from print media poses significant challenges due to limited access, high costs, and the proprietary nature of data held by newspapers or third-party vendors. \citet{gentzkow2014competition} highlighted these difficulties in their historical analysis of U.S. newspapers, where data collection involved extensive archival research and manual digitization. The scarcity of digitized archives for print newspapers contrasts sharply with the abundance of data available from digital media, creating a barrier for researchers aiming to conduct comprehensive studies on print media content and its impact.

\citet{puglisi2015empirical} discussed the obstacles in assessing media bias and content diversity due to these accessibility issues. The proprietary nature of print media archives often requires researchers to secure expensive licenses or subscriptions, limiting the scope and scale of potential studies. \citet{peiser2000cohort} noted that declining newspaper readership has led to reduced investments in archiving and data preservation, further exacerbating the problem. 
%
Advancements in technology, particularly in Optical Character Recognition (OCR) and computational methods, have begun to mitigate some of the challenges associated with analyzing print media at scale. \citet{smith2007overview} provided an overview of the Tesseract OCR engine, an open-source tool that has significantly improved the accuracy and efficiency of converting scanned print documents into machine-readable text. This development enables researchers to process large volumes of print media content, facilitating quantitative analyses that were previously impractical due to resource constraints.

In this work, we explore these dynamics in a fully automated manner by leveraging physical newspaper data to conduct a systematic analysis of the ad ecosystem and quantify the impact of governmental and corporate advertisements on editorial bias. By employing computational techniques, we aim to analyze large-scale datasets, capturing the nuances of how advertisements affect the tone and volume of media coverage in Indian newspapers. To the best of our knowledge, we are the first to conduct such an analysis using an automated mechanism, and all data and code used in this study are available \footnote{\url{https://github.com/harsha20032020/Code_and_Data_for_Analyzing_Patterns_and_Influence_of_Advertising_in_Print_Newspaper}} to facilitate reproducibility and further exploration by the research community.
Our approach extends the existing literature by offering a data-driven, scalable method to assess the role of advertiser influence across various factors such as source and entity type.

\begin{figure*}
    \centering
    \includegraphics[width=0.9\linewidth]{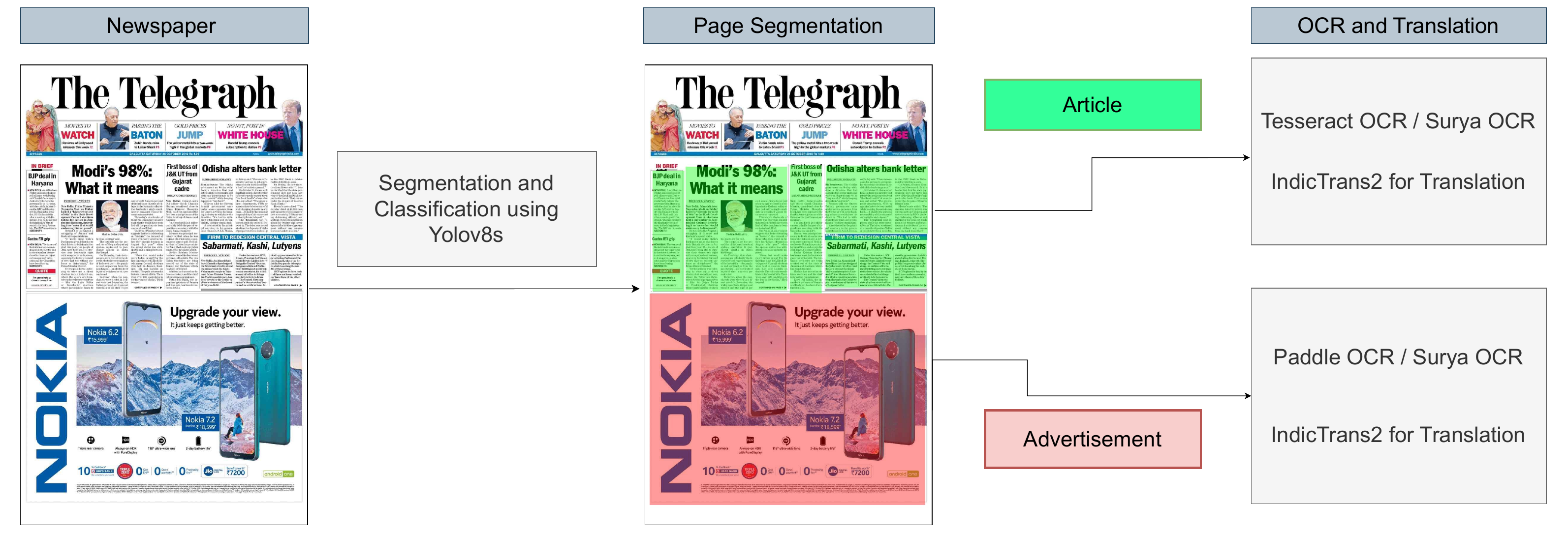}
    \caption{Processing epaper pages into textual entries across multiple sources and languages.}
    \label{fig:main_flowchart}
    \vspace{-\baselineskip}
\end{figure*}

\section{Data}
Our data collection involves scraping the archives of the print newspapers (also known as `epapers') in India.
We focus on five major news sources, which are among the country's most popular: \textit{The Times of India},
\textit{Hindustan Times},
 and \textit{The Telegraph}
(English sources), as well as \textit{Dainik Bhaskar}
(in Hindi), and \textit{Sakshi}
(in Telugu). 
Times of India is India's most circulated English newspaper; Hindustan Times is a close second, and The Telegraph is also widely read and most circulated in the state of West Bengal.
Dainik Bhaskar and Sakshi are among the most circulated Hindi and Telugu newspapers. Some of these sources are chosen because of their high circulation and consistent availability of e-papers for the time period of interest.
For each source, we crawled e-papers from the respective sites from the timelines mentioned in Table \ref{tab:newspaper_comparison}. We also collected papers from the same sources from multiple metropolitan cities to understand any regional effect.


The data collected primarily consists of page-level images. However, additional metadata was available for The Times of India and Hindustan Times. This metadata includes bounding box information for different page segments, division of articles and advertisements, and textual data corresponding to articles, though advertisements lacked associated text. This availability of metadata aided us in making a custom image segmentation model for processing other sources. Since the article text was already provided, we only needed to apply Optical Character Recognition (OCR) to the advertisements for these sources. 


\subsection{Image Segmentation Model}
To analyze the distinction between advertisements and articles in newspaper pages, we utilized data from the Hindustan Times, which had information focusing on bounding boxes that indicate whether the content was an advertisement or an article.
This data comprised 1,024 images, divided into 851 images for training, 151 images for testing, and 22 images for validation. After preprocessing the dataset with standard image preprocessing (fixing orientation and resizing), we finetuned YOLOv8s \cite{yolov8_ultralytics} model for this task, which is particularly well-suited for object detection tasks like identifying object bounding boxes. The model's performance metrics are as follows: a mean Average Precision (mAP) of 96.8\%, a Precision of 86.9\%, and a Recall of 88.8\%. These results demonstrate the model's high accuracy and robustness in distinguishing between advertisements and articles in our dataset and segmenting the epaper page properly into respective article/ad segments.
Despite the availability of additional data, we observed diminishing returns when incorporating a larger dataset (around 10k samples) into the training process.
Performance metrics for larger dataset can be found in Appendix \ref{appendix_seg_model}.

\subsection{OCR and Translation}
Once the Image Segmentation model processes the images, for each individual segment of the image identified as an advertisement or article, we apply a combination of OCR models based on the language and content type. For English-language articles, we use Tesseract OCR \cite{TessOverview} because its page segmentation modes effectively retrieves text in the manner of reading order which makes it suitable for structured article content where text flow is crucial,  moreover, it achieves an error rate of less than 5\%, ensuring high accuracy in text retrieval.

However, Tesseract OCR faces significant challenges in English-language advertisements due to irregular layouts, diverse fonts, and random text arrangements, which reduce Tesseract's performance. To overcome this, we use PaddleOCR~\cite{paddlepaddleocr}, which maximizes text extraction from more complex layouts, such as advertisements, when the reading order is not a priority. Since the reading sequence in advertisements is generally less structured, PaddleOCR's robust extraction capabilities make it the ideal tool for this task.

For content in Indic languages (Hindi and Telugu in our dataset), we employ Surya OCR \cite{suryaocrcite}, as both PaddleOCR and TesseractOCR perform poorly with these languages and Surya OCR's performance being close to 1\% error rate. Although Surya does not extract text in reading order, we address this limitation by leveraging its ordering and segmentation capabilities. We first use Surya to segment the text components in reading order, generate a new image based on these segments, and then apply the OCR process. Appendix \ref{appendix_surya} shows a detailed example of this process. Although Surya OCR does not perform as well on advertisement data compared to articles, it remains the most effective solution currently available for processing Indic-language text \cite{limitationssurya}.

After extracting the Indic language text using Surya OCR, we translate it into English using the IndicTrans2~\cite{gala2023indictrans} model, which is capable of translating both Hindi and Telugu, ensuring that we can convert the regional language content into English for further analysis. Since our study does not primarily analyze the content of the articles and only uses keywords to filter different advertisers/content, the minor errors introduced by translation do not affect the study.

\section{Analysis of the Print Ad Ecosystem}
\label{sec:exploratory_analysis}

Our dataset offers a unique opportunity to analyze the nature of advertisements in print newspapers at an unprecedented scale. The methodology we developed is highly scalable and applicable to any newspaper, regardless of language or nation, potentially providing much-needed transparency in the print advertising landscape. 

We focus on differentiating between government and corporate advertising expenditures.\footnote{Unless explicitly specified, all the plots and analyses refer to the combined dataset; separate plots are available in the Appendix} These two categories of advertisers roughly make up 50\% of both the volume and spending on advertising. Advertising on the first, third, and last pages of a physical newspaper holds significant importance due to higher visibility and reader engagement~\cite{MediaScience2024, RJI2024}. These pages are considered premium spots and command higher advertising prices.

\noindent\textbf{Keyword Identification}.
For identifying articles and ads related to governments, we first identify articles related to government corruption using a manually curated set of keywords such as ``bribe,'' ``scam,'' ``corrupt,'' and other relevant terms commonly found in corruption-related articles. Government advertisements are classified by detecting keywords like ``government,'' ``state,'' and ``tender'' in the ad text. In this study, ``government'' is used specifically to refer to state and national institutions rather than local political entities or parties. The complete set of keywords used to identify both government articles and ads is provided in Appendix~\ref{govt_keywords}. Our keyword-based approach demonstrates strong performance, achieving an F1 score of 0.94 and an accuracy of 0.94 when validated against 100 advertisement samples, confirming the completeness of our keyword selection.
For the corporate analysis, we select a subset of the most prominent advertisers reported in TAM Media Research India's quarterly reports.\footnote{\url{https://tamindia.com/}} These companies represent key sectors such as Fast-Moving Consumer Goods (FMCG), Automobile, Technology, Education, and other major industries. Keywords associated with each company are derived from popular products and common terms used to refer to that company. A comprehensive list of the companies and their corresponding keywords can be found in Appendix~\ref{company_keywords}. Our methodology assumes that any potential misclassifications at the keyword identification stage are randomly distributed across page numbers and page areas, which are the only two factors that could potentially influence our regression analyses and graphical representations, thus minimizing systematic bias in our findings.
Using these keywords, we extract relevant articles and advertisements for each entity, enabling a focused analysis of their advertising patterns.

\subsection{Who is advertising?}
\begin{figure}[h]
    \centering
    \includegraphics[width=0.9\linewidth]{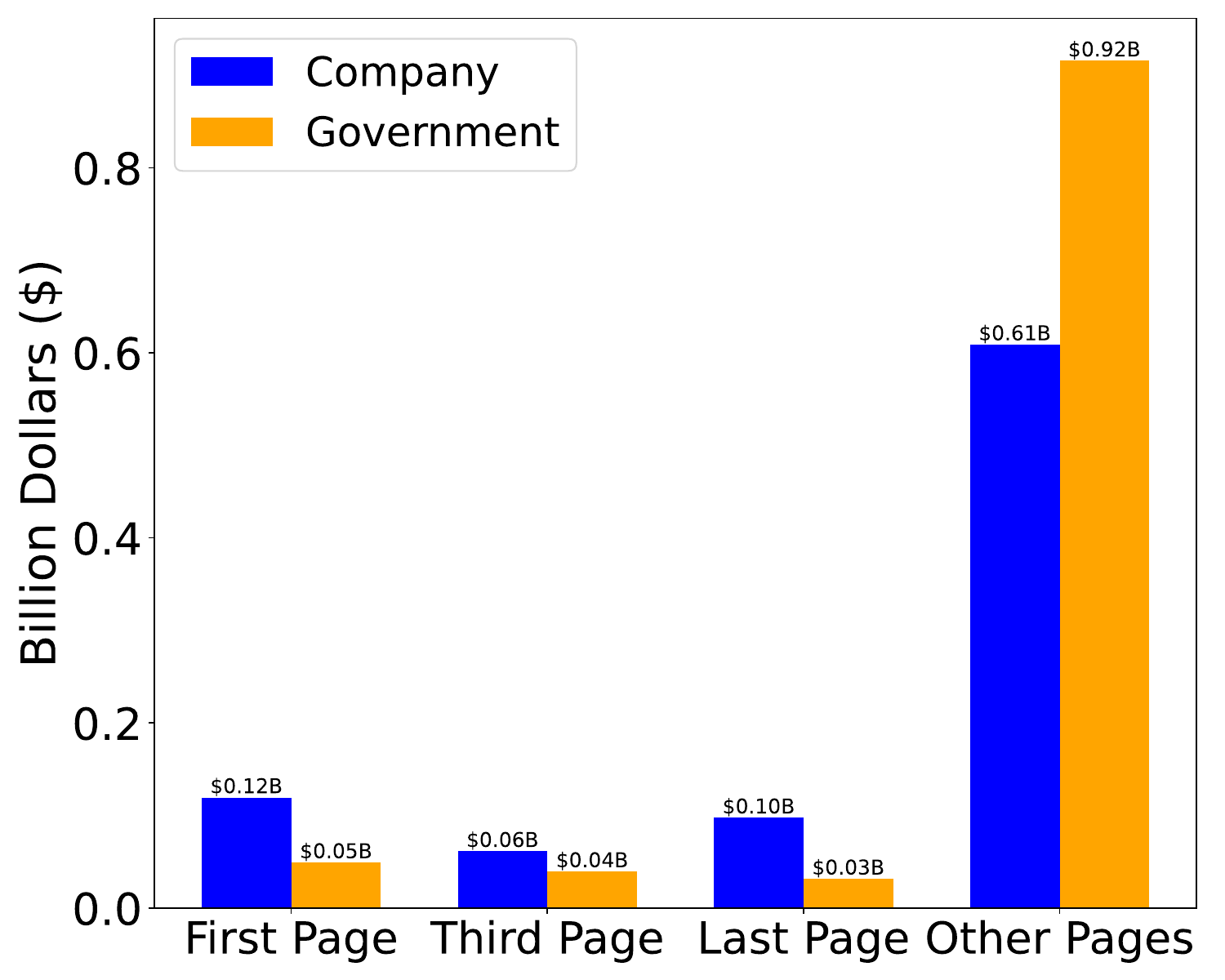}
    \caption{Spending by Companies and Government.}
    \label{fig:spending_absolute}
    \vspace{-\baselineskip}
\end{figure}

Figure~\ref{fig:spending_absolute} illustrates the absolute spending by government and companies. The data indicates that government entities are significant contributors to print advertising, surpassing corporate spending in total expenditure.
Companies account for \$890 million in advertising expenditure in our dataset, while government spending reaches \$1.02 billion—with the majority of newspaper ads being government-placed. These cost estimates are derived by multiplying the ad's area with the cost per square centimeter that accounts for multiple factors, including each advertisement’s page area, placement location, page number, and the specific newspaper source, along with the corresponding cost per square centimeter. A detailed breakdown of these variations and their associated rates is provided in Table \ref{tab:page_rates}.

\begin{figure*}[h] 
    \centering
    \begin{subfigure}{0.24\textwidth}
        \includegraphics[width=\linewidth]{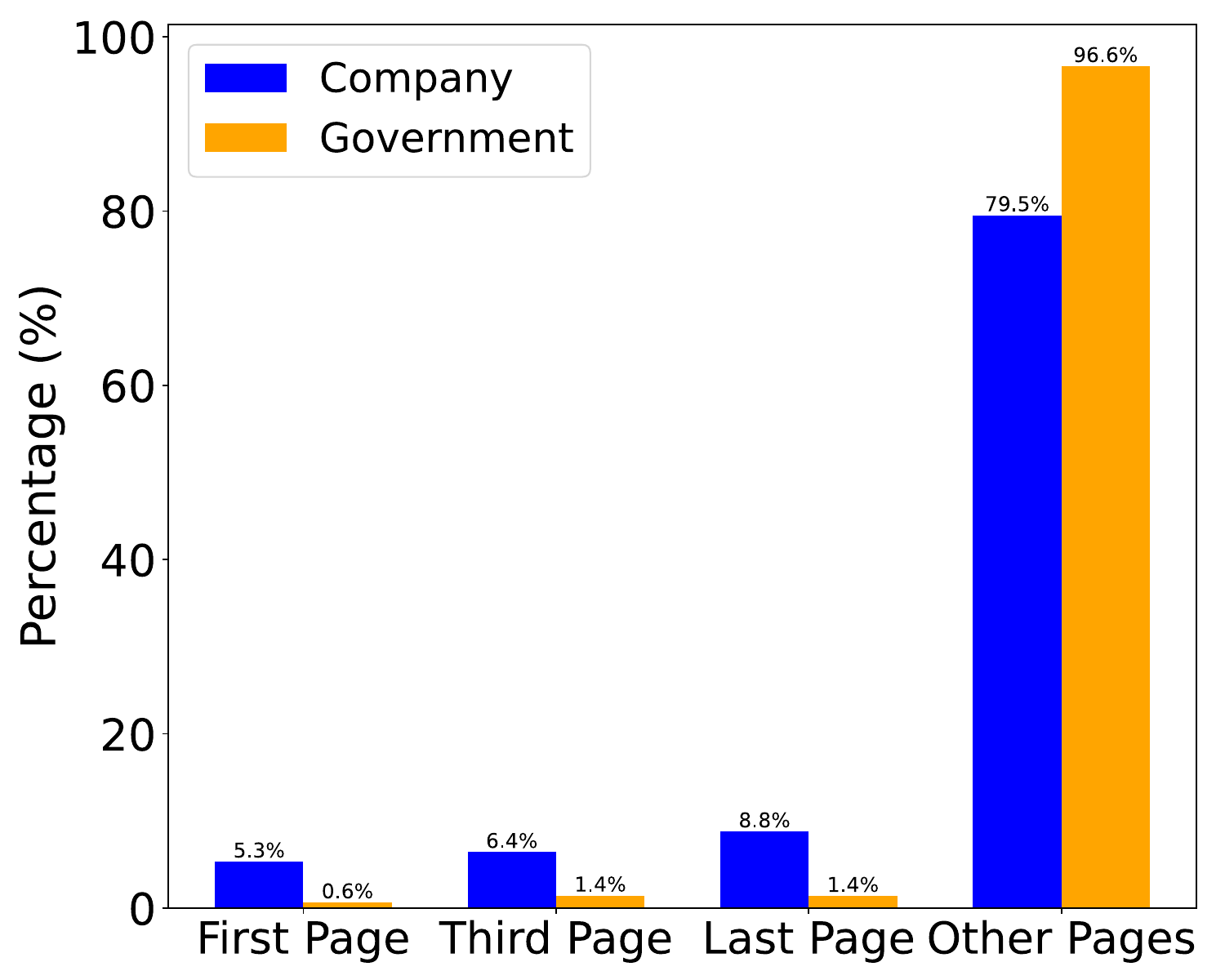}
        \caption{Percentage of ads on a specific page}
        \label{fig:image1}
    \end{subfigure}
    \hfill 
    \begin{subfigure}{0.24\textwidth}
        \includegraphics[width=\linewidth]{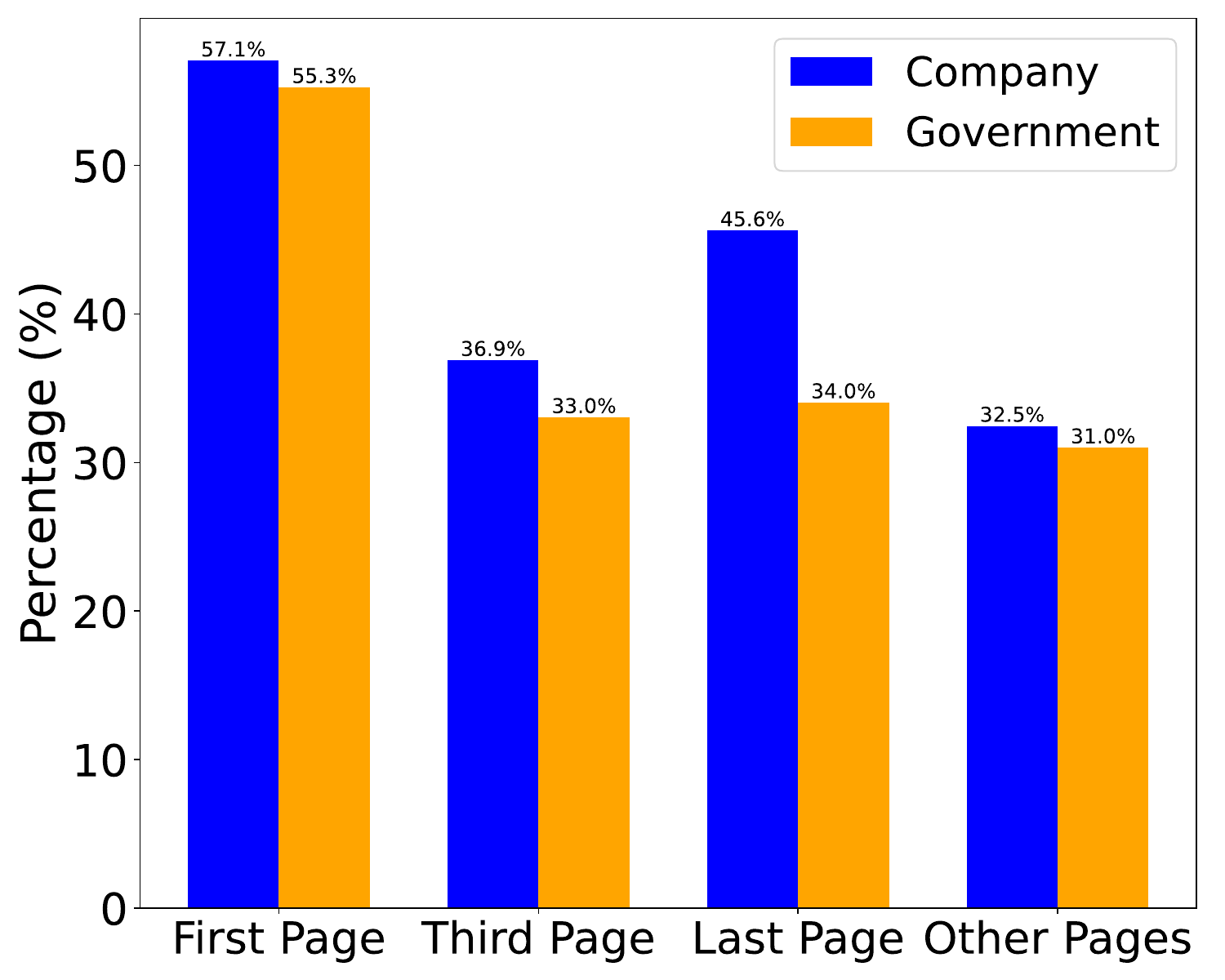}
        \caption{Percentage of page area occupied by ads on average}
        \label{fig:image2}
    \end{subfigure}
    \hfill
    \begin{subfigure}{0.24\textwidth}
        \includegraphics[width=\linewidth]{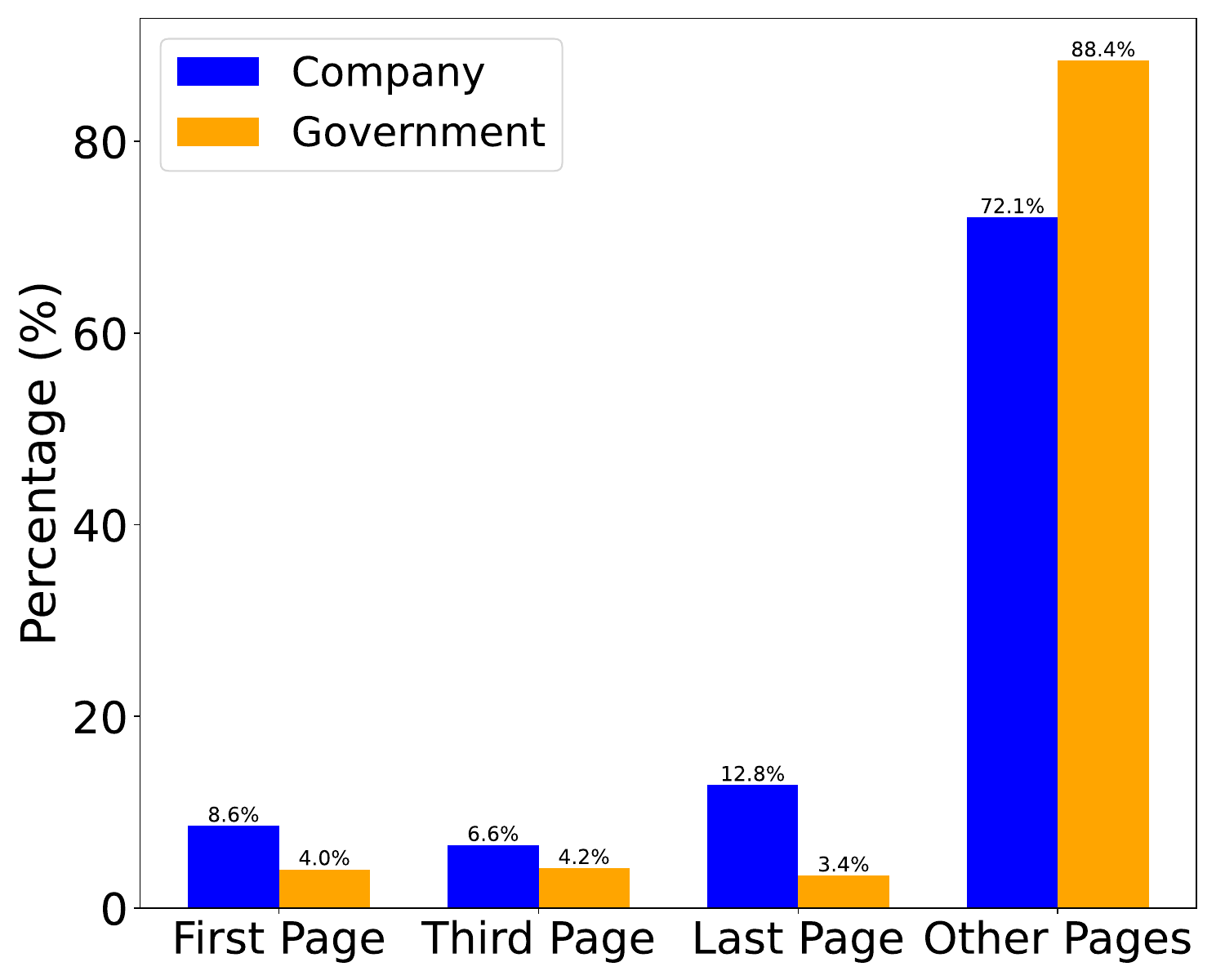}
        \caption{Percentage of page area in comparison to total area}
        \label{fig:image3}
    \end{subfigure}
    \hfill
    \begin{subfigure}{0.24\textwidth}
        \includegraphics[width=\linewidth]{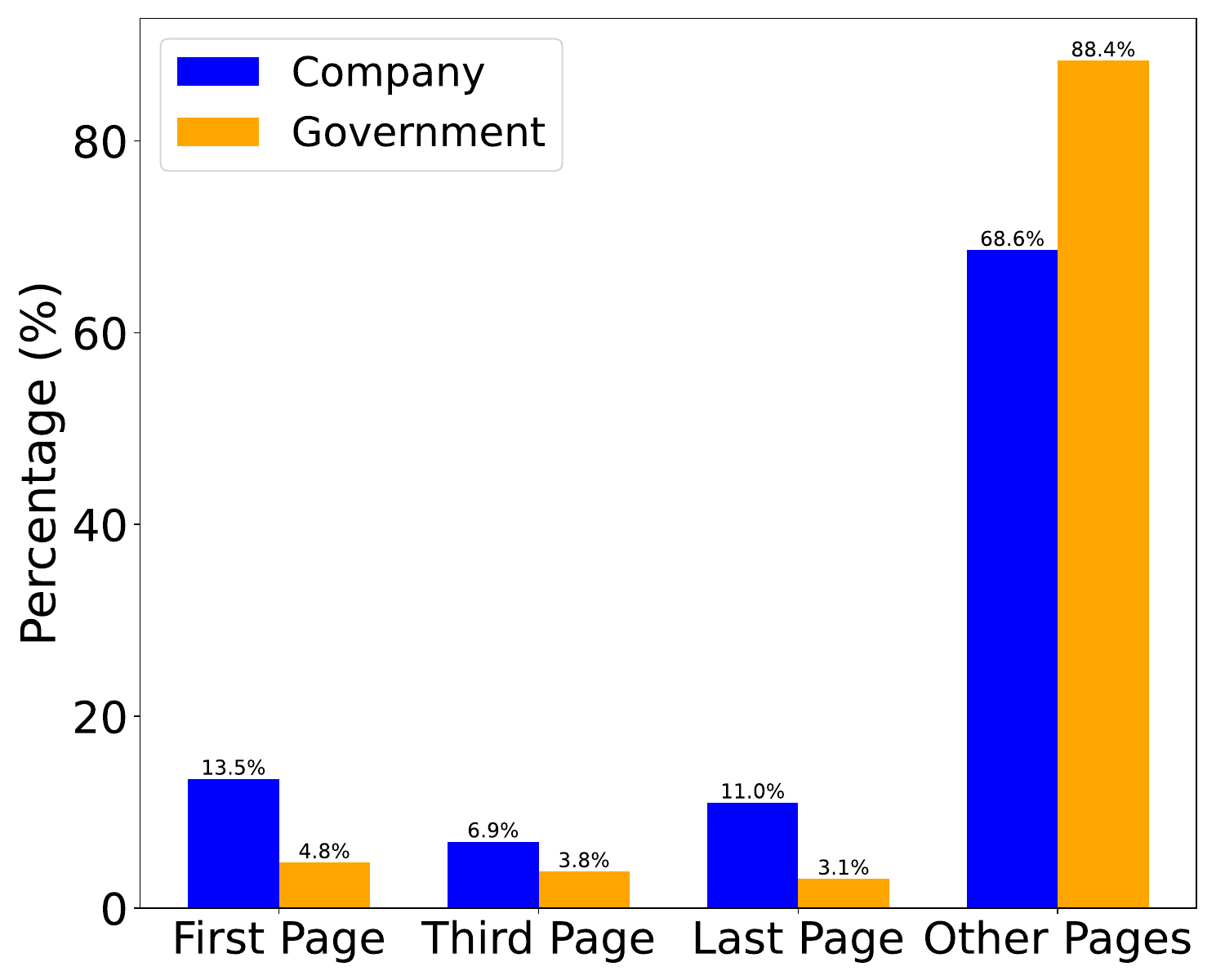}
        \caption{Percentage of ad expenditure compared to total ad spending}
        \label{fig:image4}
    \end{subfigure}
    \caption{Where are ads being placed?}
    \vspace{-\baselineskip}
    \label{fig:descriptives}
\end{figure*}
\subsection{Where are the ads being placed?}

Following industry conventions (detailed in Appendix Section \ref{appendix_price_page}), we categorize ads based on their placement: \textbf{Premium Pages:} 1st, 3rd, and last pages. \textbf{Other Pages:} All remaining pages.
To quantify the advertising presence and expenditure, we compute four main variables for each entity (government or companies):

\begin{enumerate} \item The percentage of all ads provided by the entity that appear on a specific page of the newspaper. \item  The average percentage of the page area covered by the entity's ads when they appear on the specified page. \item The percentage of the total specified page area occupied by the entity's ads, calculated across all ads by the entity. This provides an overall measure of the entity's visibility on the specified page, considering both the frequency and size of ad appearances. \item The entity's total spending on newspaper ads, quantifying the entity's financial investment in their print advertising efforts. \end{enumerate}

Our analysis yields the following results, as shown in Figures~\ref{fig:descriptives}(a--d):

(i) \textbf{Ad Placement Frequency (Figure~\ref{fig:descriptives}a):}  A majority of government ads are not on the first, third, or last pages. Specifically, 88.4\% of government advertising is found on other pages. Conversely, companies have a significantly higher fraction of ads on these premium pages, with 27.9\% of their ad area appearing there.

(ii) \textbf{Ad Size on Premium Pages (Figure~\ref{fig:descriptives}b):} When ads appear on the front page, government ads occupy an average of 55.3\% of the page area, while company ads occupy 57.1\%. This indicates that although the government advertises less frequently on premium pages, the size of their ads is comparable to that of companies when they do.

(iii) \textbf{Total Page Area Occupied (Figure~\ref{fig:descriptives}c):} After normalizing by the total area of ads from either companies or the government, front-page ads contribute 8.56\% of the total area for company advertisements. The combined 1st, 3rd, and last pages account for 27.9\% of the company ad area, highlighting their emphasis on premium placements.
In contrast, only 11.6\% of government advertising is found on these pages.

(iv) \textbf{Advertising Expenditure Distribution (Figure~\ref{fig:descriptives}d):} By calculating the sum of ad costs for specific pages and dividing by the total ad expenditure for company ads, we find that companies allocate 31.6\% of their total ad expenditure to just the 1st, 3rd, and last pages. In contrast, government spending on advertisements is more evenly distributed across all pages, without disproportionate emphasis on specific pages.

\subsection{How do they advertise?}

To understand the methods and strategies employed by different advertisers, we analyze the sizes and placements of advertisements, which reveal insights into the types of ads provided, the approximate amounts spent, and the fraction of the page area they cover. By computing the size of the ads, we can discern patterns in advertising behavior between government entities and companies.

\begin{figure}[h]
    \centering
    \includegraphics[width=0.8\linewidth]{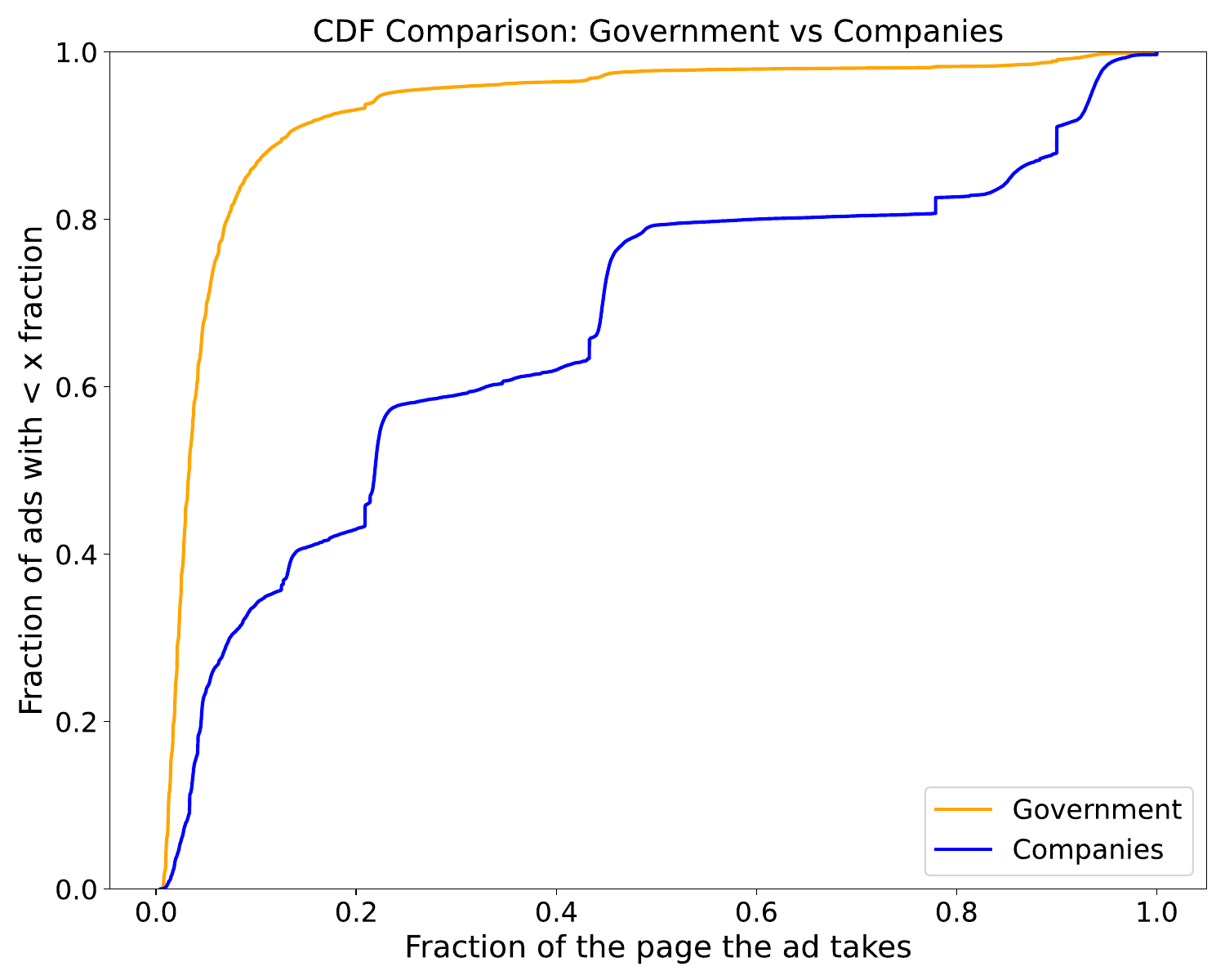}
    \caption{CDF of the area fraction of the ads showing government advertisers' strong preference for smaller ads (under 10\%) and corporate advertisers' distinct focus on quarter-page, half-page, and full-page ads.}
    \label{fig:cdf}
    \vspace{-\baselineskip}
\end{figure}

Figure~\ref{fig:cdf} presents the cumulative distribution function (CDF) of advertisement sizes for both government and corporate advertisers. The data reveals a stark contrast between the two groups. Approximately 85\% of the ads placed by government entities are small, occupying less than 10\% of a page. Full-page ads by the government are exceedingly rare, constituting only about 1\% of their total advertisements. 
%
In contrast, corporate advertisers display a different pattern. The CDF for companies shows distinct jumps at the 25\%, 50\%, and 100\% marks, corresponding to quarter-page, half-page, and full-page ads, respectively. 

Further analysis uncovers additional observations regarding advertising strategies:
We document an odd-page bias in advertisement placement. Advertisers prefer to place ads on odd-numbered pages, which are typically the right-hand pages in a newspaper layout. This preference may be an artifact of the way newspapers are read, as right-hand pages are more immediately visible when flipping through the pages. Our data supports this bias, with corporate ads appearing more frequently on odd pages (42,724 instances) compared to even pages (29,393 instances). This finding aligns with industry practices that consider odd-numbered pages as premium spots due to higher reader engagement.

Our dataset also allows us to compute the targeting priorities of various companies. For example, companies like Amazon and Patanjali have been primarily advertising in Hindi-language newspapers, as illustrated in Appendix Figure~\ref{fig:companies_percentage_ads}. 
We also observe interesting differences in advertising approaches across national English-language newspapers and regional publications. Companies tend to place more full-page ads in English-language newspapers, as shown in Appendix Figure~\ref{fig:company_ads_across_papers}. 
%
In general, most advertisements are heavily concentrated at smaller sizes, covering between 5\% to 10\% of a page, as depicted in  Figure~\ref{fig:total_ads_across_pages} (Appendix). 
Additionally, clear trends emerge when analyzing advertisement sizes across different industries and sectors. Educational institutions predominantly use full-page ads, as shown in Appendix Figure~\ref{fig:education_ads_across_papers1}
. 
In contrast, insurance companies typically opt for half-page ads (Appendix Figure~\ref{fig:insurance_ads_across_papers}), whereas technology companies exhibit clustering around quarter-page, half-page, and full-page ads (Appendix Figure~\ref{fig:technology_ads_across_papers}).
\subsection{When are the ads being placed?}
\begin{figure}[ht]
    \centering
    \includegraphics[width=\linewidth]{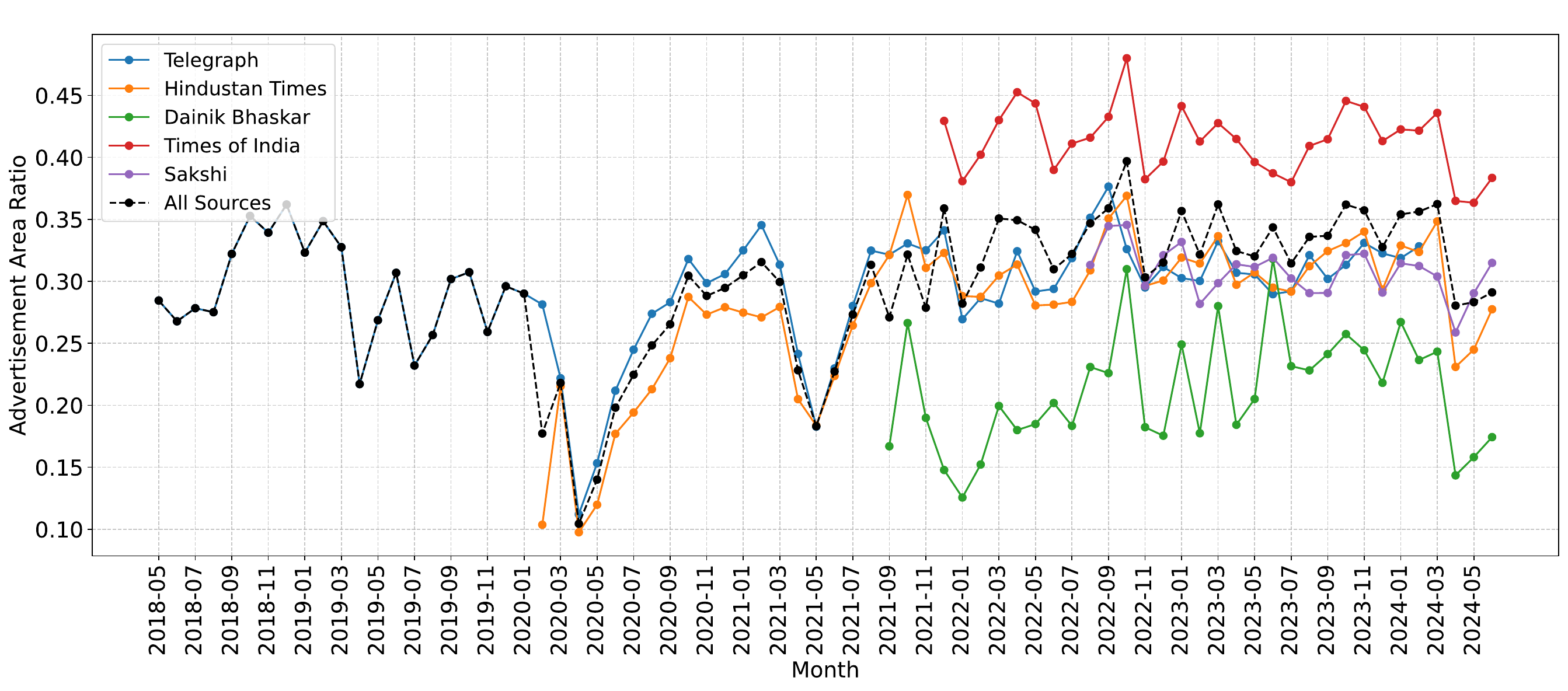}
    \caption{Monthly Advertisement Area Ratio by Source with Aggregate.}
    \label{fig:timeseries}
    \vspace{-\baselineskip}
\end{figure}

To understand the temporal dynamics of advertising in print newspapers, we conducted an analysis of time-based trends, as depicted in Figure~\ref{fig:timeseries}. This figure illustrates the monthly average percentage of newspaper space occupied by advertisements from May 2018 to May 2024.
Overall, the data reveals that advertisements consistently occupy approximately 35\% of the total newspaper area throughout the years.

However, significant fluctuations are observed during certain periods:
In early 2020, coinciding with the onset of the COVID-19 pandemic, there is a marked decline in the percentage of newspaper space devoted to advertisements. This drop reaches its lowest point in April and May 2020, where advertising space plummets to around 10\%. The decline can be attributed to the economic uncertainty and reduced consumer activity during nationwide lockdowns, prompting advertisers to cut back on spending.
A similar but smaller drop can also be found in June and July 2021, which aligns with the second wave of COVID-19 in India. This period was characterized by severe health impacts and renewed economic disruptions, leading to a temporary reduction in advertising activity.

As the largest English daily, with a readership of over 15 million, Times of India (TOI) consistently has a higher percentage of its space occupied by advertisements, averaging around 40\%. This indicates a strong preference among advertisers for TOI, likely due to its wide reach among English-speaking and urban audiences.
In contrast, Dainik Bhaskar, a leading Hindi-language daily, exhibits an average advertising space of approximately 20\%, nearly half that of TOI. 

Despite short-term fluctuations, the overall stability in the proportion of advertising space over the 6 year period suggests that print media remains a vital platform for advertisers. While digital media has grown substantially, print advertising continues to play a significant role, especially in reaching certain demographics and regions where newspapers are a primary source of information.




\subsection{What are the ads about?}
To understand the content focus of the advertisements, we conducted topic classification on the ad texts. We used an off-the-shelf topic classification model from Hugging Face,\footnote{\url{https://huggingface.co/dstefa/roberta-base_topic_classification_nyt_news}} specifically the RoBERTa-base model fine-tuned on a New York Times dataset. 

\begin{figure}[ht]
    \centering
    \includegraphics[width=\linewidth]{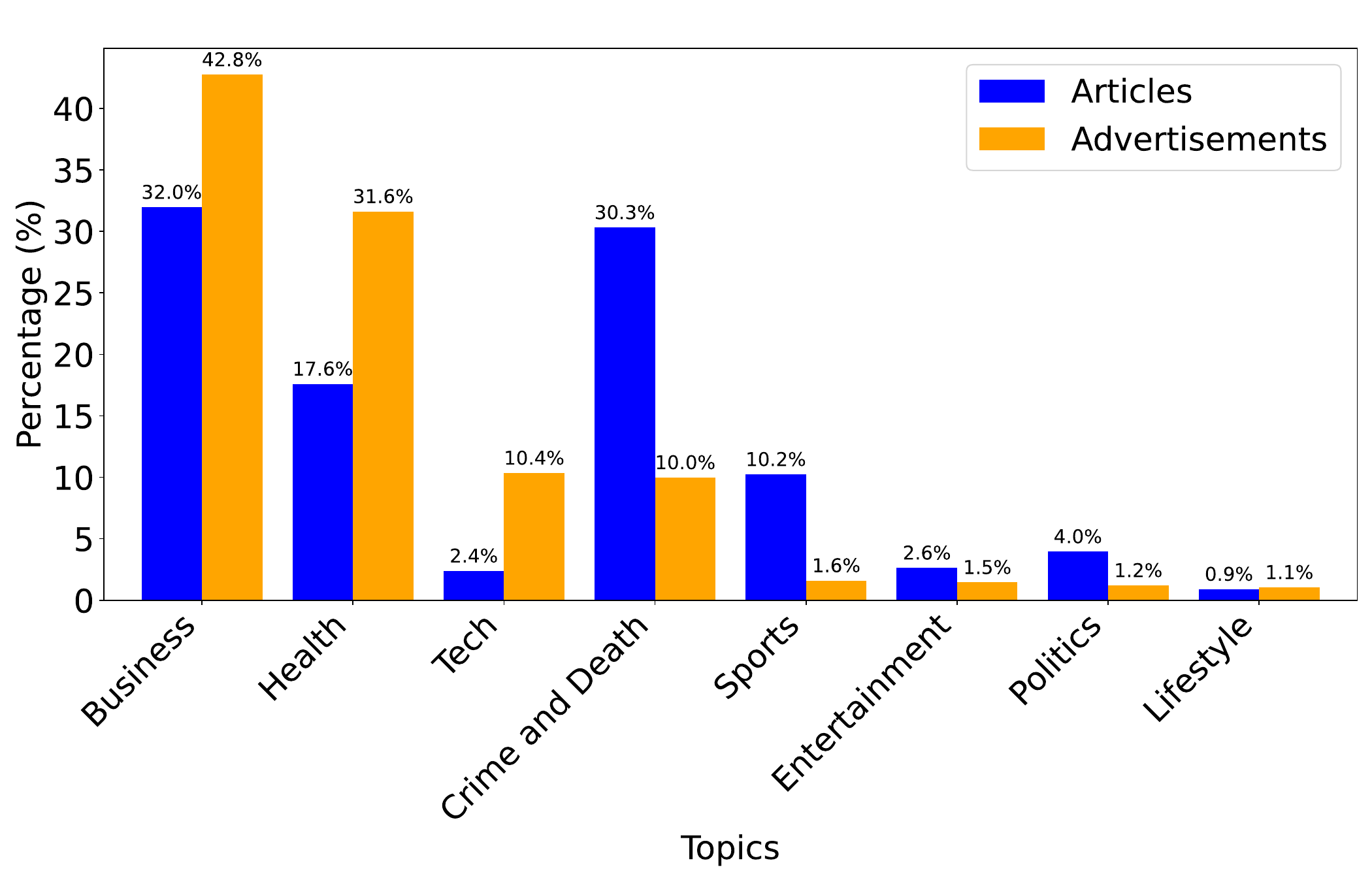}
    \caption{Topics covered in ads and in the article content.}
    \label{fig:topics_ads}
    \vspace{-\baselineskip}
\end{figure}

This model provides classification into eight classes. Evaluation of the classification model can be found in Appendix~\ref{sec:exploratory_analysis_plots_appendix}. Figure~
\ref{fig:topics_ads} presents the distribution of topics in the advertisements compared to the articles. Our analysis reveals several interesting observations. Business, health, and technology ads are significantly higher in proportion than articles in these topics. 
Conversely, topics such as crime and death, which are more prevalent in articles (including obituaries in the case of ads), are less represented in advertisements. 

\section{Do advertisers get positive coverage?}
Building upon the data and insights presented in the previous sections, we address a pivotal research question: \textit{Does advertising in a newspaper influence the coverage that advertisers receive in the news content?} Specifically, we aim to explore how advertising revenue from both corporate and government entities may affect the portrayal of news, regarding content bias and the extent of coverage. Leveraging our extensive dataset, we conduct a thorough analysis to examine this potential relationship.

Our investigation focuses on two primary dimensions: the \textbf{coverage} of entities, measured by the number of articles mentioning them, and the \textbf{sentiment} of that coverage, determined through sentiment of the articles. By examining these factors in relation to the advertising expenditure of the entities, we seek to understand whether there is a correlation between advertising spending and favorable media representation.

Inspired by previous literature that explores similar questions in specific contexts --such as the influence of car industry advertising on news coverage~\cite{ditella} or the reporting of government corruption~\cite{mentionedtwice}-- our approach builds upon established models from economics. We introduce measurable proxies for both advertising spending and media coverage to facilitate a quantitative analysis.

For advertising spending, we employ the \textit{weighted ad ratio} as a proxy. This ratio quantifies the prominence and investment in advertisements by considering both the size of the ads and their placement within the newspaper. The ratio is calculated using the following formula:

\[
\mathrm{Weighted\ Ad\ Ratio} = \mathrm{Scaling\ Factor} \cdot \frac{\mathrm{Ad\ Area}}{\mathrm{Page\ Area}}
\]

The scaling factor is computed to indicate the relative importance and cost associated with the placement of an ad. Advertising rates vary depending on factors such as the page number, edition, and city. We obtained ad rates for various editions and page numbers for all the newspapers in our dataset to compute these scaling factors. For instance, if the base rate per square centimeter for an ad in The Times of India is $\$x$, and an ad on the front page costs $\$y$, the scaling factor for front-page ads is calculated as $y/x$. The scaling factors used in our analysis are based on actual rates from the newspapers, as detailed in Table~\ref{tab:page_rates}.
A comprehensive description of the costs and scaling factors can be found in Appendix~\ref{appendix_price_page}.\footnote{While we could directly use the estimated costs from Section~\ref{sec:exploratory_analysis_plots_appendix}, variations in costs across advertisers and the approximate nature of these estimates led us to use the area of the ads, adjusted by scaling factors, as a more robust measure.}

By incorporating the scaling factor and normalizing by the page area, the weighted ad ratio allows for meaningful comparisons across different pages and publications, despite differences in page sizes and advertising costs. This normalization ensures that our analysis accounts for the relative prominence and investment in advertisements, rather than absolute measures that may be skewed by publication-specific characteristics.

Our regression analysis examines the relationship between the sentiment and volume of media coverage over a given time period and the weighted ad ratio of advertisements placed by the entities. By focusing on the difference in sentiment and coverage volume, we aim to explore whether increased advertising expenditure correlates with more favorable or extensive media portrayal, without relying on direct financial data that may be proprietary or unavailable.

Sentiment analysis of the articles was conducted using the model from ~\cite{camacho-collados-etal-2022-tweetnlp}, one of the most widely adopted and downloaded models on Hugging Face providing sentiment classification of the content: -1, 0, or 1 corresponding to negative, neutral, or positive sentiment.
Since we saw in Section~\ref{sec:exploratory_analysis} that government spending and types of ads are very different from other corporates, we will do the analysis separately.

\begin{table*}
\centering
\caption{Comprehensive Panel Regression Results: The Impact of Ad Page Percentage on Sentiment and Article Count.}
\label{tab:comprehensive_results}
\phantomsection\label{tab:regression_results_govt}
\phantomsection\label{tab:regression_govt_counts}
\phantomsection\label{tab:regression_company_sentiment}
\phantomsection\label{tab:regression_company_count}
\begin{tabular}{llccccc}
\hline
\textbf{Focus} & \textbf{Dependent Variable} & \textbf{Model} & \textbf{Ad Page \% Coef.} & \textbf{Fixed Effects} & \textbf{Entities} & \textbf{Periods} \\
\hline
Government & Total Sentiment & (1) & -0.0730*** & None & 5 & 75 \\
 &  & (2) & -0.0115*** & Source & 5 & 75 \\
 &  & (3) & -0.0709*** & Time & 5 & 75 \\
 &  & (4) & -0.0084* & Source + Time & 5 & 75 \\
 & Count of Articles & (1) & 0.5810*** & None & 5 & 75 \\
 &  & (2) & 0.1041* & Source & 5 & 75 \\
 &  & (3) & 0.5337*** & Time & 5 & 75 \\
 &  & (4) & 0.0214 & Source + Time & 5 & 75 \\
\hline
Companies & Total Sentiment & (1) & 0.0189*** & None & 155 & 72 \\
 &  & (2) & 0.0139*** & Newspaper × Company & 155 & 72 \\
 &  & (3) & 0.0170*** & Time & 155 & 72 \\
 &  & (4) & 0.0137*** & Newspaper × Company + Time & 155 & 72 \\
 & Count of Articles & (1) & 0.4360*** & None & 149 & 70 \\
 &  & (2) & 0.2225*** & Newspaper × Company & 149 & 70 \\
 &  & (3) & 0.4366*** & Time & 149 & 70 \\
 &  & (4) & 0.2232*** & Newspaper × Company + Time & 149 & 70 \\
\hline
\multicolumn{7}{l}{\textit{Note: Standard errors (omitted for brevity) are clustered by entity. Significance levels: * p$<$0.1, ** p$<$0.05, *** p$<$0.01}} \\
\end{tabular}
\end{table*}

\subsection{Regression setup}

We employ a panel regression approach to understand the correlation between advertisement and news coverage, given that our dataset is structured with observations across multiple entities (such as news outlets and regions) over time.

The correlation between the advertising space and article sentiment (or coverage) is then analyzed using a panel regression model. Specifically, we regress the total sentiment of the articles (or coverage) on the weighted ad ratio occupied by the company's advertisements while incorporating fixed effects for media-company combinations and time \cite{beattie2017}.
The regression follows the structure:

\begin{align*}
\mathrm{Total\ Sentiment}_{mnt} &= \alpha + \beta \cdot \mathrm{Weighted\ Ad\ Ratio}_{mnt} \\
&\quad + \gamma_{mn} + \delta_t + \epsilon_{mnt}
\end{align*}

where $\alpha$ represents the intercept, serving as the baseline level of the dependent variable. The coefficient $\beta$ quantifies the effect of the percentage of page area occupied by company $n$'s advertisements in newspaper $m$ during time period $t$, denoted as $\text{area}_{mnt}$. The term $\gamma_{mn}$ captures the media-company fixed effects, accounting for unobserved heterogeneity between different newspapers and companies. Time-fixed effects are incorporated through $\delta_t$, which accounts for variations over daily, weekly, or monthly time periods. Finally, $\epsilon_{mnt}$ is the error term that captures the unexplained variation in sentiment.

For companies, the $n$ refers to each company. For government, we consider all government ads as a single entity. We perform two types of regressions -- one for sentiment and one for coverage. This formulation allows us to isolate the specific relationship between the percentage of advertisement area and the sentiment of articles while controlling for both unobserved differences across media-company pairs and time-based fluctuations and capture the relationship between advertisement percentage and sentiment. To further examine if popularity correlates with sentiment and count of companies involved in the regression. We also consider a regression with an added popularity term (likely influencing how positively the coverage is framed) and with fixed effects on company and time. 
To obtain the popularity of a company, we used Google trends data and computed trend values for popularity for all terms corresponding to a company.

\subsection{Findings}
For government, across most models, Table \ref{tab:regression_results_govt}, there is a negative correlation between weighted ad ratio and sentiment, indicating that an increase in the percentage of page area occupied by advertisements is associated with more negative sentiment in articles. This may be because, unlike companies that can selectively choose which newspapers to advertise in, government-related works (e.g. tenders and contracts) and announcements must be published in all major newspapers by law. This requirement removes the incentive for newspapers to tailor positive coverage in exchange for government advertisements, resulting in a more neutral or negative tone.
For companies, Table \ref{tab:regression_company_sentiment} demonstrates a robust and positive correlation between \(\mathrm{Weighted\ Ad\ Ratio}\)
 and both \(\mathrm{Sentiment}\) and total \(\mathrm{Coverage}\). In all model specifications, the coefficient for ad page percentage remains statistically significant, indicating that increased advertising correlates with more favorable sentiment (and coverage) in news articles. Even when controlling for newspaper × company and time-fixed effects, the positive impact of advertising persists.

Results indicate strong and significant effects, showing that a 1\% increase in a company's weighted ad ratio leads to an average increase of 0.0189 units in the total sentiment score. Given the narrow sentiment scale, which ranges from -1 to 1, this increase of 0.0189 units is substantial.




While considering the popularity of company terms from Google trends, we find a surprising relation. In Appendix Table 
\ref{tab:regression_popular_sentiment}, we notice that including the popularity term into the regression, the correlation between sentiment and advertising decreases and is non-significant when accounting for company fixed effects. This suggests that the influence of ad spending on sentiment may be less straightforward, where popularity might play a more complex role. However, overall, neither ad spending nor popularity shows consistent significance on sentiment across all specifications when company and time effects are included.

Even after accounting for popularity, weighted ad ratio remains highly significant across all models, indicating a robust correlation between ad page percentage and the count of articles (Appendix Table \ref{tab:regression_popular_counts}). This mirrors the results from Table~\ref{tab:regression_company_count}, where ad page percentage alone was a strong predictor of article volume, suggesting that higher ad spending continues to drive media coverage, even when company popularity is taken into account. Notably, the effect of popularity itself is inconsistent across models, with a negative correlation in some models and non-significant results in others, highlighting that ad spending is a more reliable driver of media attention.




\section{Conclusion}

In this paper, we introduced a unique pipeline for collecting and analyzing advertisements in print newspapers at scale. Our methodology is highly scalable and applicable to any newspaper, irrespective of language or region, opening new avenues for large-scale research on the effects of advertisements in print media. By open-sourcing our code and dataset, we aim to enable researchers and practitioners to further explore the intricacies of print advertising and its impact on media content.

Our comprehensive analysis across multiple languages and newspapers revealed significant insights into the scale and nature of both government and corporate advertising. By examining who advertises, where, when, how, and what they advertise, we uncovered distinct patterns in advertising strategies. Government entities are major advertisers, with a wide distribution of smaller ads across various pages, while companies tend to invest in premium placements and larger ad formats to maximize visibility.

Through our regression analysis, we found a clear, statistically significant correlation between corporate advertising expenditure and both the volume and tone of the media coverage they receive. This suggests that higher advertising spending by companies may correlate with more favorable and extensive coverage in the news content. In contrast, the correlation for government entities was less pronounced, possibly due to the obligatory nature of many government ads, which are legally mandated and broadly disseminated, making it challenging to influence coverage at scale.
These findings raise important questions about media bias in favor of advertisers and its potential effects on public discourse, informed citizenship, and democratic processes. The influence of advertising revenue on editorial content underscores the need for transparency and ethical considerations within the media industry.

Our dataset remains highly underutilized, offering rich opportunities for further research. Notably, we possess the actual text of the articles that appeared adjacent to the advertisements, enabling studies on contextual influence and placement effects. Additionally, our dataset serves as a valuable resource for comparative analyses with digital media, allowing for explorations of cross-media advertising strategies and their implications.
\bibliographystyle{ACM-Reference-Format}
\bibliography{biblio}

\clearpage

\input{appendix}

\end{document}

%% file: appendix.tex
\appendix


\section{Segmentation Model and Outputs}
\label{appendix_seg_model}
The image segmentation model processes newspapers' pages and identifies segments that are either articles or advertisements. One such example is given below

\begin{figure}[H]
    \includegraphics[width=0.8\linewidth]{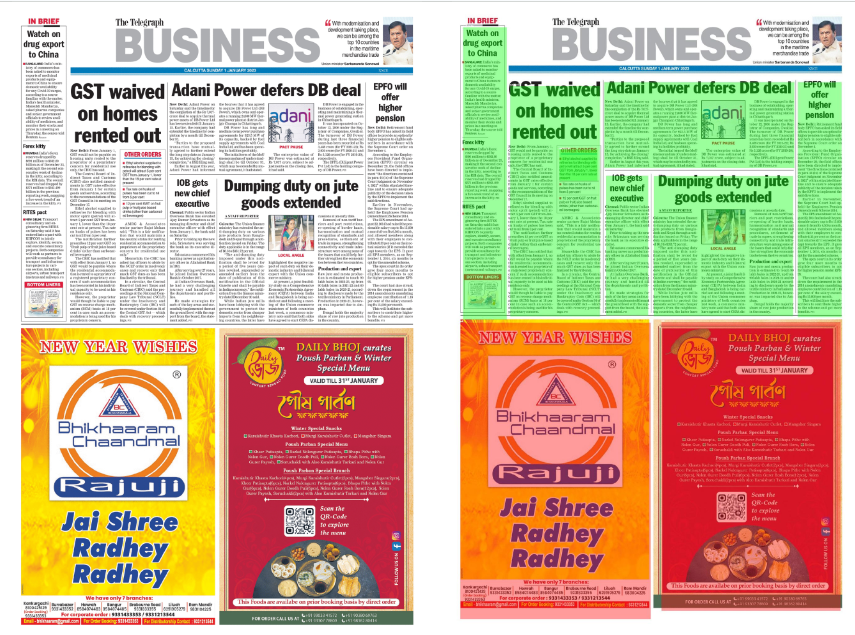}
    \caption{Newspaper page before and after segmentation}
    \label{fig:enter-label}
\end{figure}

The segments are then cropped and processed through the rest of the pipeline. The segmentation model processes the images with confidence and IOU thresholds of 0.1, and then the identified regions are cropped with a margin of 10 pixels. The model and code used for inference will be made public. \\

We also used 1024 pages for training the model despite having access to larger training datasets, primarily since the performance degraded with larger datasets. Given the importance of a balanced approach in identifying both articles and advertisements accurately, we selected the first model for further analysis due to its superior mAP, which provides a better overall measure of detection performance across all classes. 

\begin{table}[ht]
\centering
\caption{Results for ad detection based on training data size}
\label{tab:results_part2}
\begin{tabular}{l|l|l|l}
\hline
\textbf{Dataset Size} & \textbf{mAP}    & \textbf{Precision} & \textbf{Recall} \\ \hline
1024         & 96.8\% & 86.9\%    & 88.8\% \\ \hline
9329         & 95.3\% & 83.5\%    & 97.8\% \\ \hline
\end{tabular}
\end{table}




\section{Exploratory analysis plots}
\label{sec:exploratory_analysis_plots_appendix}
Approximately 2.4 million articles and advertisements were identified across the analyzed news sources. A detailed breakdown of the number of articles and advertisements per source is provided in Table \ref{tab:newspaper_comparison}. 
To understand the distribution of topics covered in the articles and advertisements, we employed a topic classification model that was fine-tuned using a dataset from The New York Times\footnote{\url{https://huggingface.co/dstefa/roberta-base_topic_classification_nyt_news}} with an accuracy and F1 score of 0.91. This model allowed us to categorize the content into predefined topics, providing a clearer understanding of the thematic focus within articles.

\begin{figure}[H]
    \centering
    \includegraphics[width=1\linewidth]{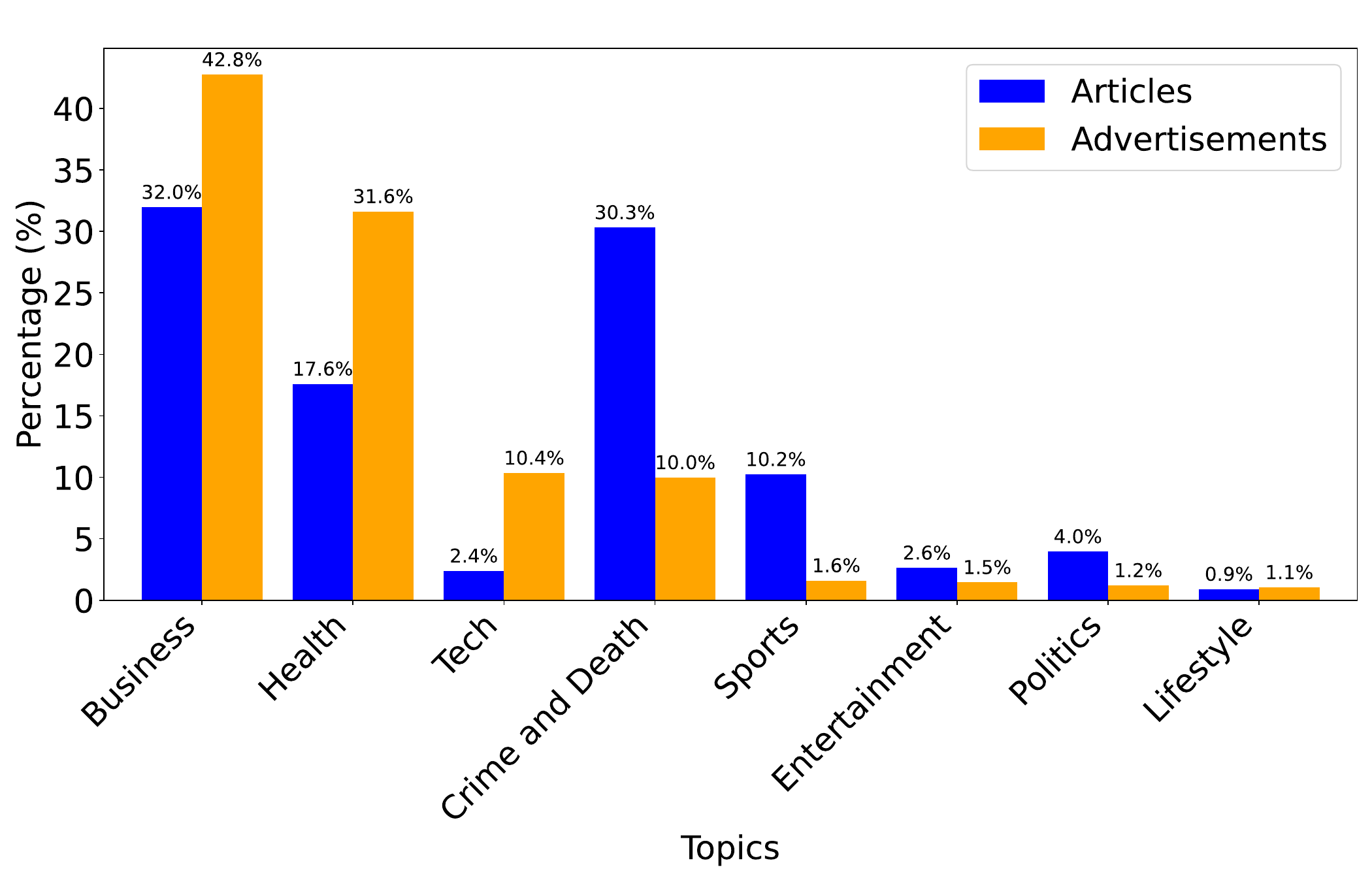}
    \caption{Distribution of Topics in Articles and Advertisements}
    \label{fig:dist_articles}
\end{figure}
\vspace{-0.3cm} 
\noindent

\begin{figure}[h]
    \centering
    \includegraphics[width=1\linewidth]{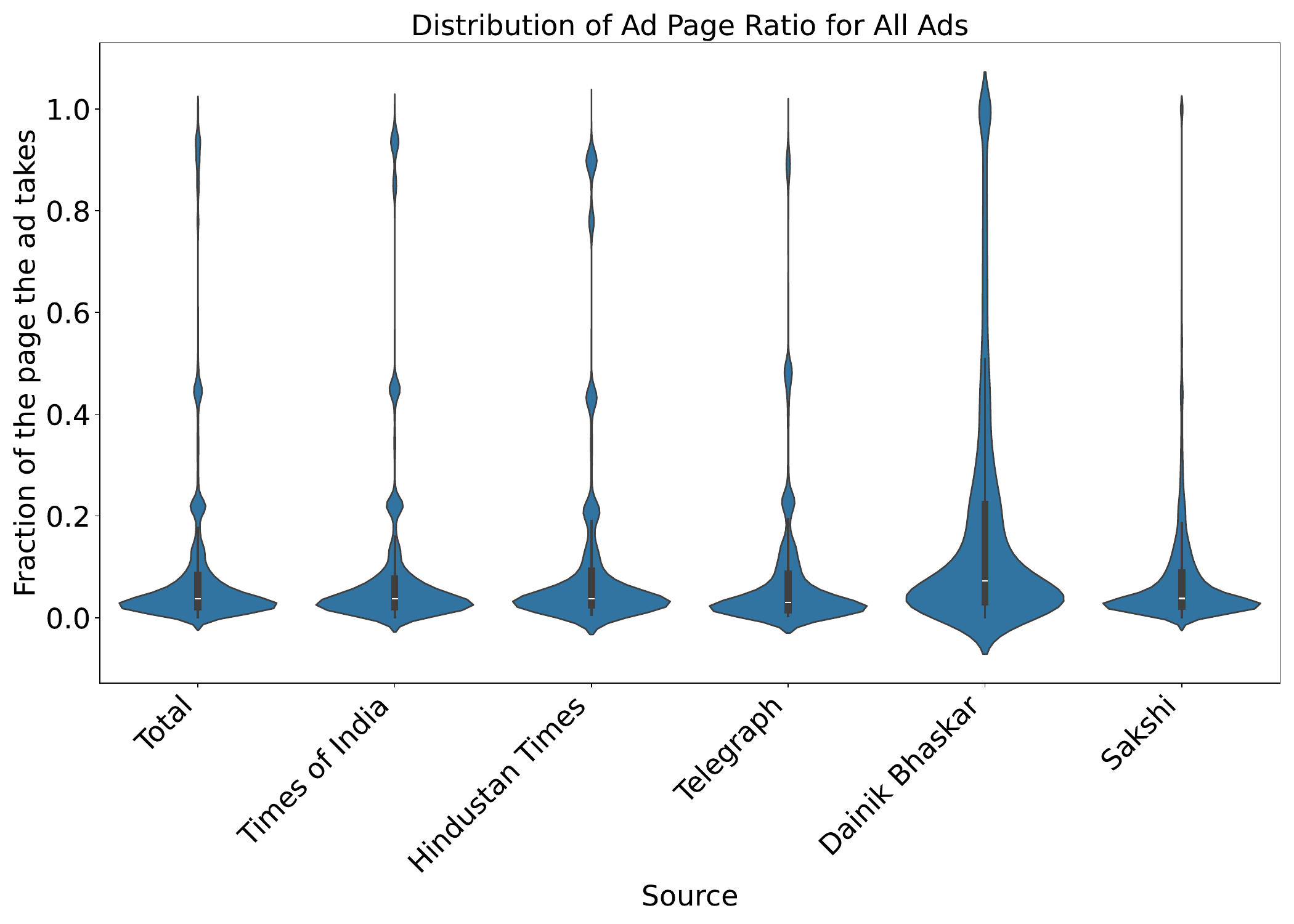}
    \caption{All ads across papers.}
    \label{fig:total_ads_across_pages}
\end{figure}

Figure~\ref{fig:total_ads_across_pages} shows the distribution of ad ratios across all papers. Most ads are small, occupying less than 10\% of the space.

\begin{figure}[H]
    \centering
    \includegraphics[width=\linewidth]{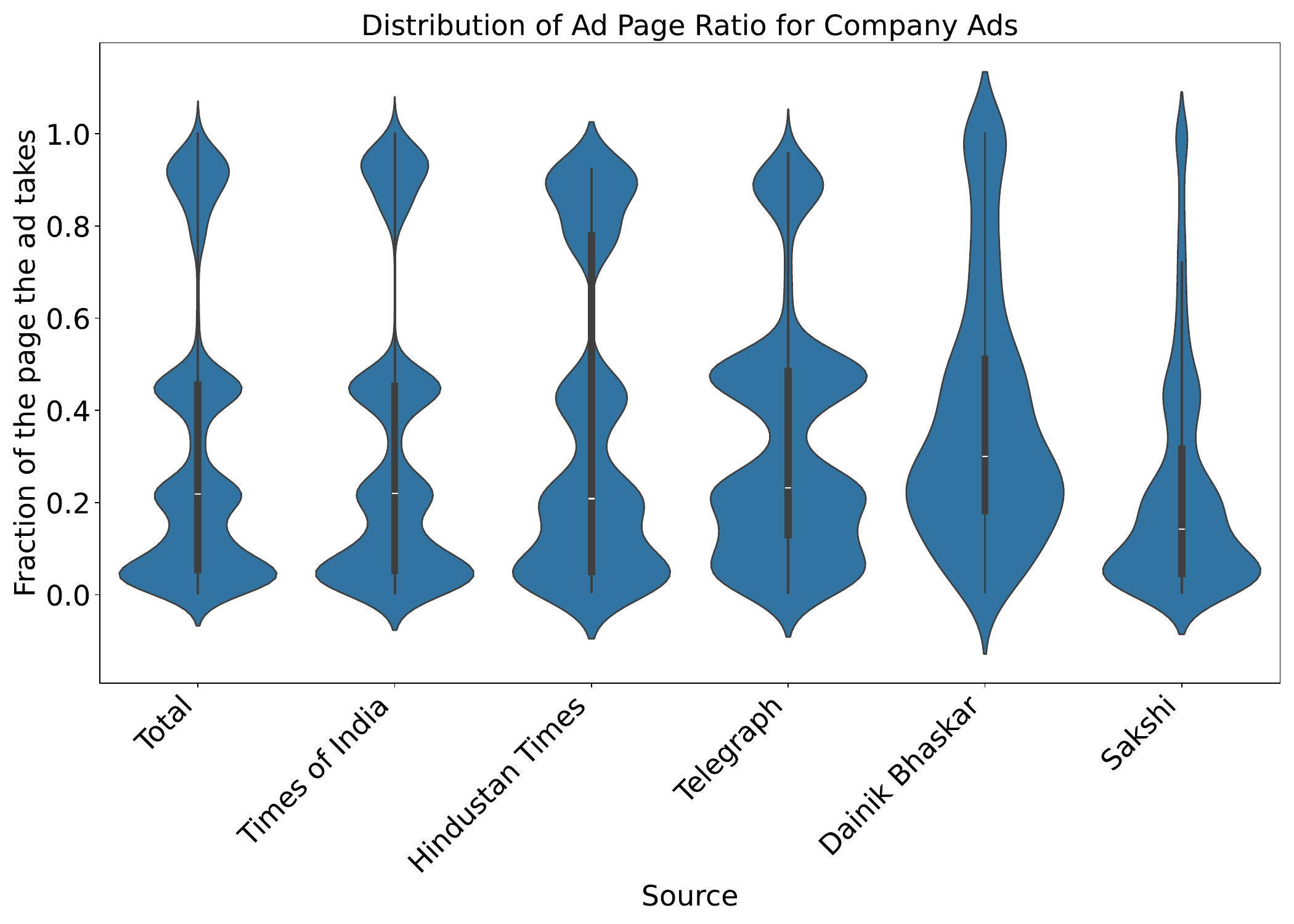}
    \caption{Company ads across papers}
\label{fig:company_ads_across_papers}
\end{figure}
Figure~\ref{fig:company_ads_across_papers} shows the distribution of company ad ratios across the 5 papers we study. Companies typically give larger ads. We can see that there is significant distribution around quarter, half, and full-page advertisements.

\begin{figure}[H]
    \centering
    \includegraphics[width=\linewidth]{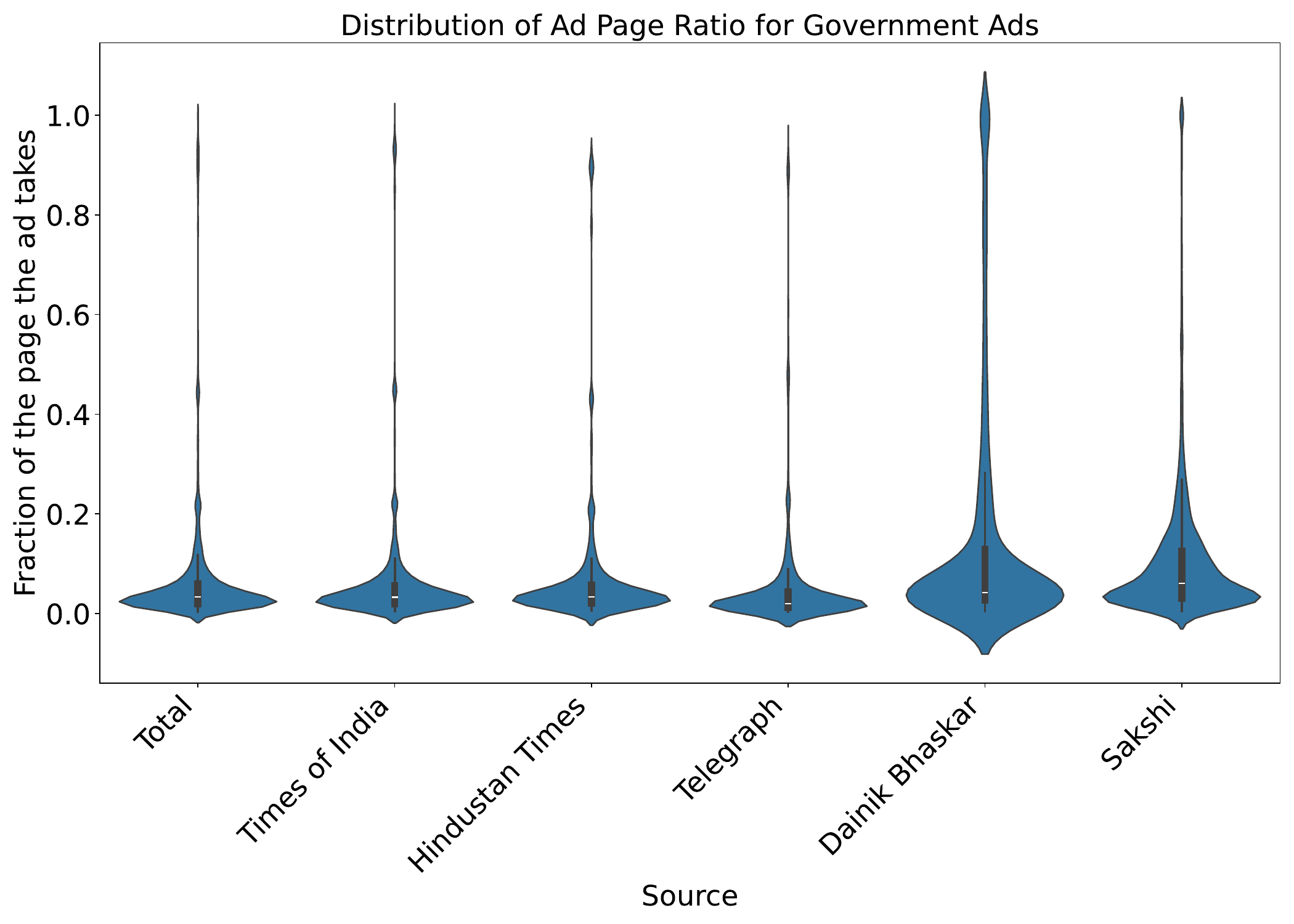}
    \caption{Government ads across papers}
    \label{fig:govt_ads_across_papers}
\end{figure}

Figure~\ref{fig:govt_ads_across_papers} shows the ads by governments. We can see that across sources, the area page ratio is small, indicating that most ads are small in comparison to the page.

\begin{figure}
    \centering
    \includegraphics[width=\linewidth]{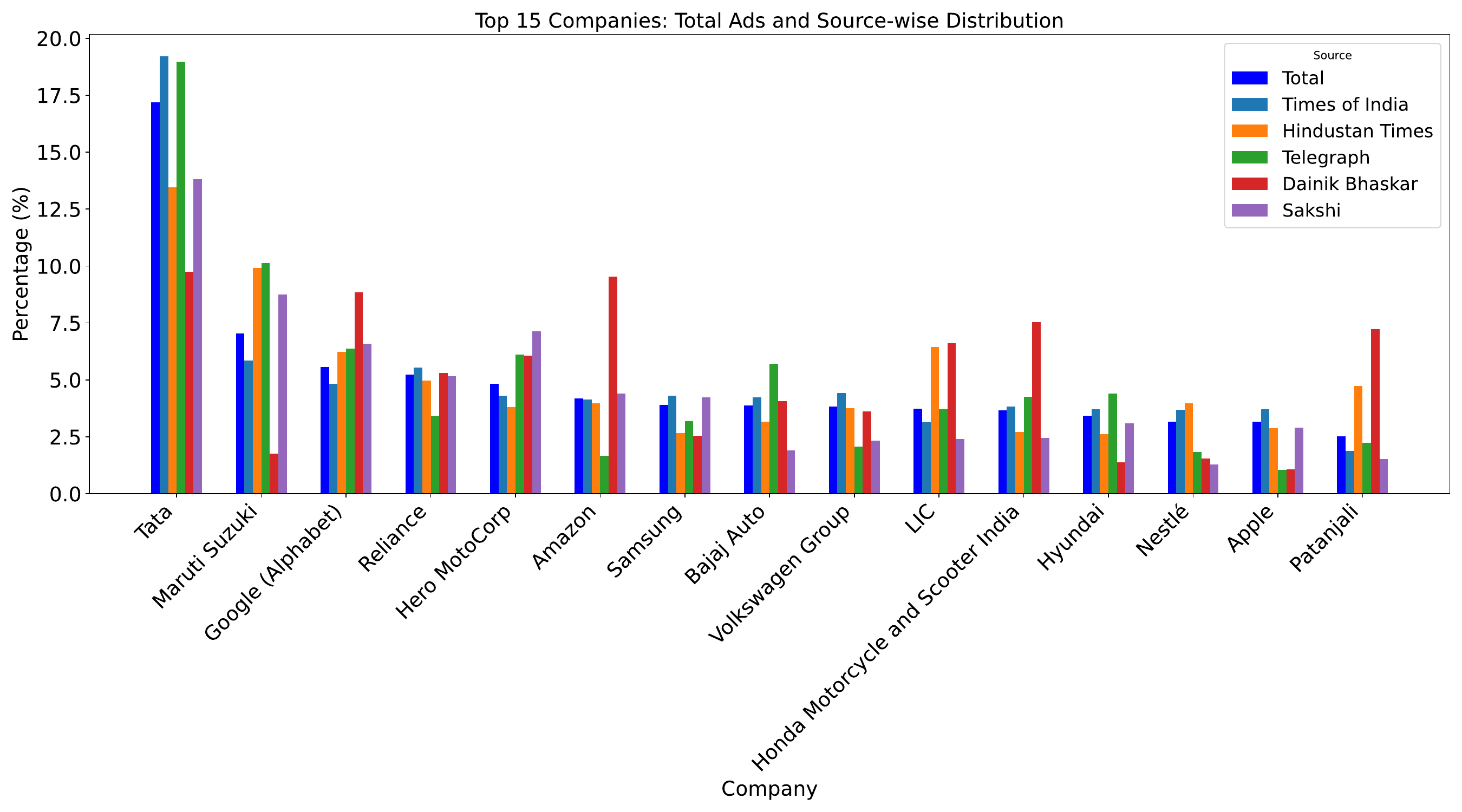}
    \caption{Percentage of ads provided by the top 15 advertisers in the 5 papers in our dataset. 
    }
    \label{fig:companies_percentage_ads}
\end{figure}

Figure \ref{fig:companies_percentage_ads} presents the distribution of the number of ads among the top 15 companies, broken down by contributions from various newspapers. Companies like Tata and Maruti Suzuki have a broad presence across multiple sources, while others focus their spending more selectively. This chart offers insight into the target audience for their industry.

\begin{figure}[H]
    \centering
    \includegraphics[width=\linewidth]{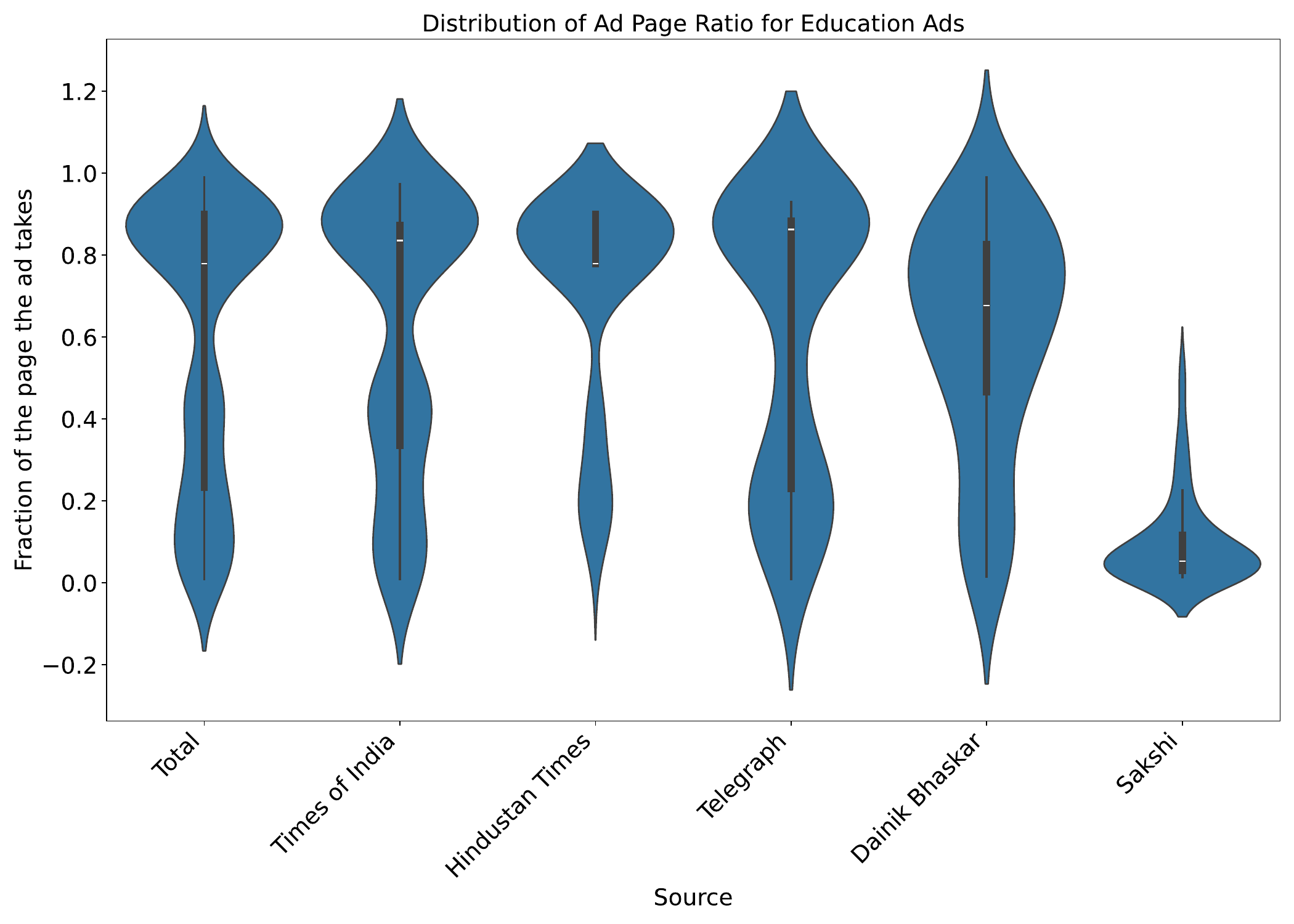}
    \caption{Education ads across papers - typically, education ads are mostly full page, except on Sakshi.}
    \label{fig:education_ads_across_papers1}
\end{figure}

\begin{figure}[H]
    \centering
    \includegraphics[width=\linewidth]{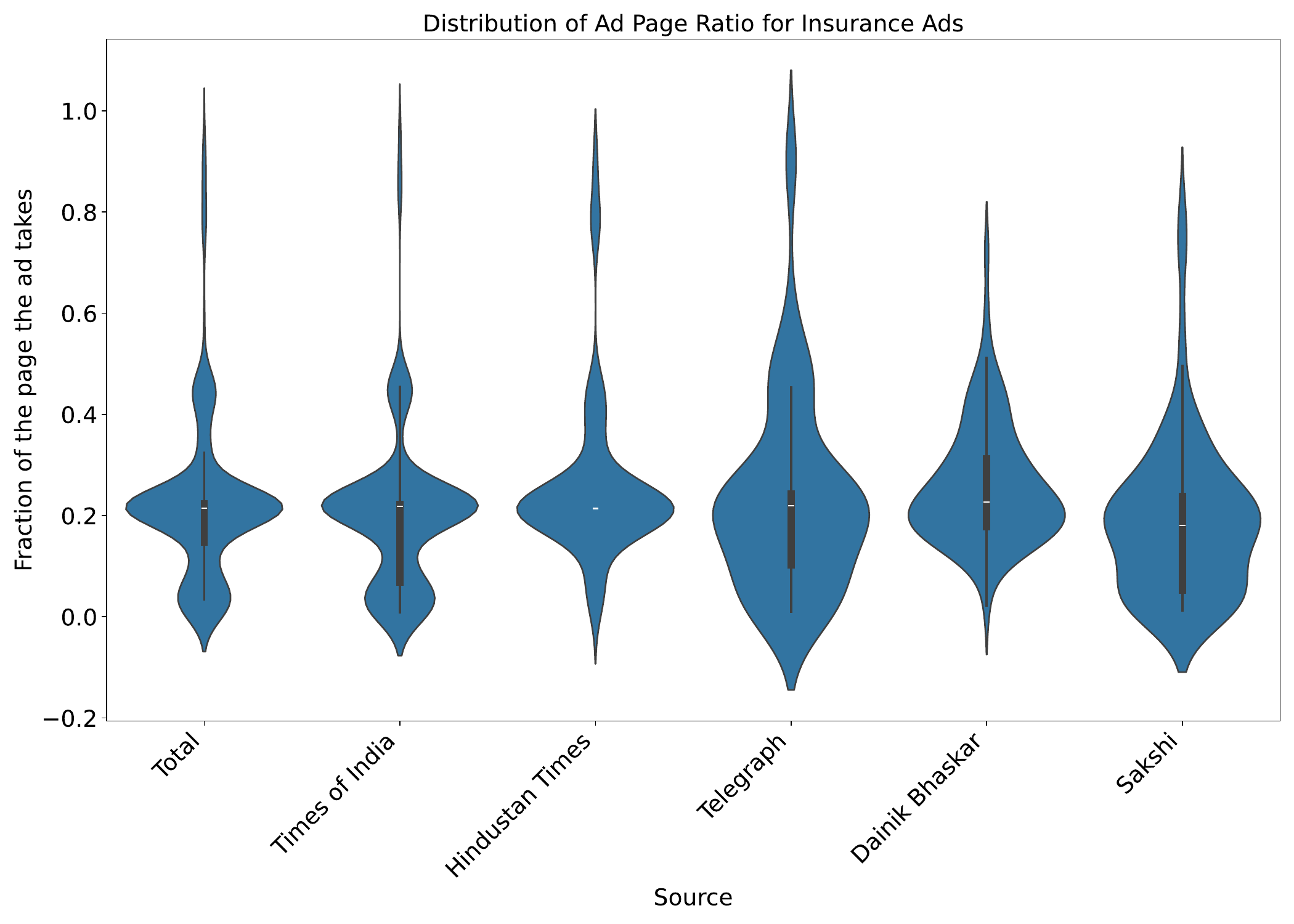}
    \caption{Insurance ads across papers - insurance ads are typically half-page}
    \label{fig:insurance_ads_across_papers}
\end{figure}

Certain sectors specifically give certain types of ads. e.g., education sector typically gives larger full-page ads (figure~\ref{fig:education_ads_across_papers1}. Insurance typically gives half-page ads (figure~\ref{fig:insurance_ads_across_papers})

\begin{figure}[H]
    \centering
    \includegraphics[width=\linewidth]{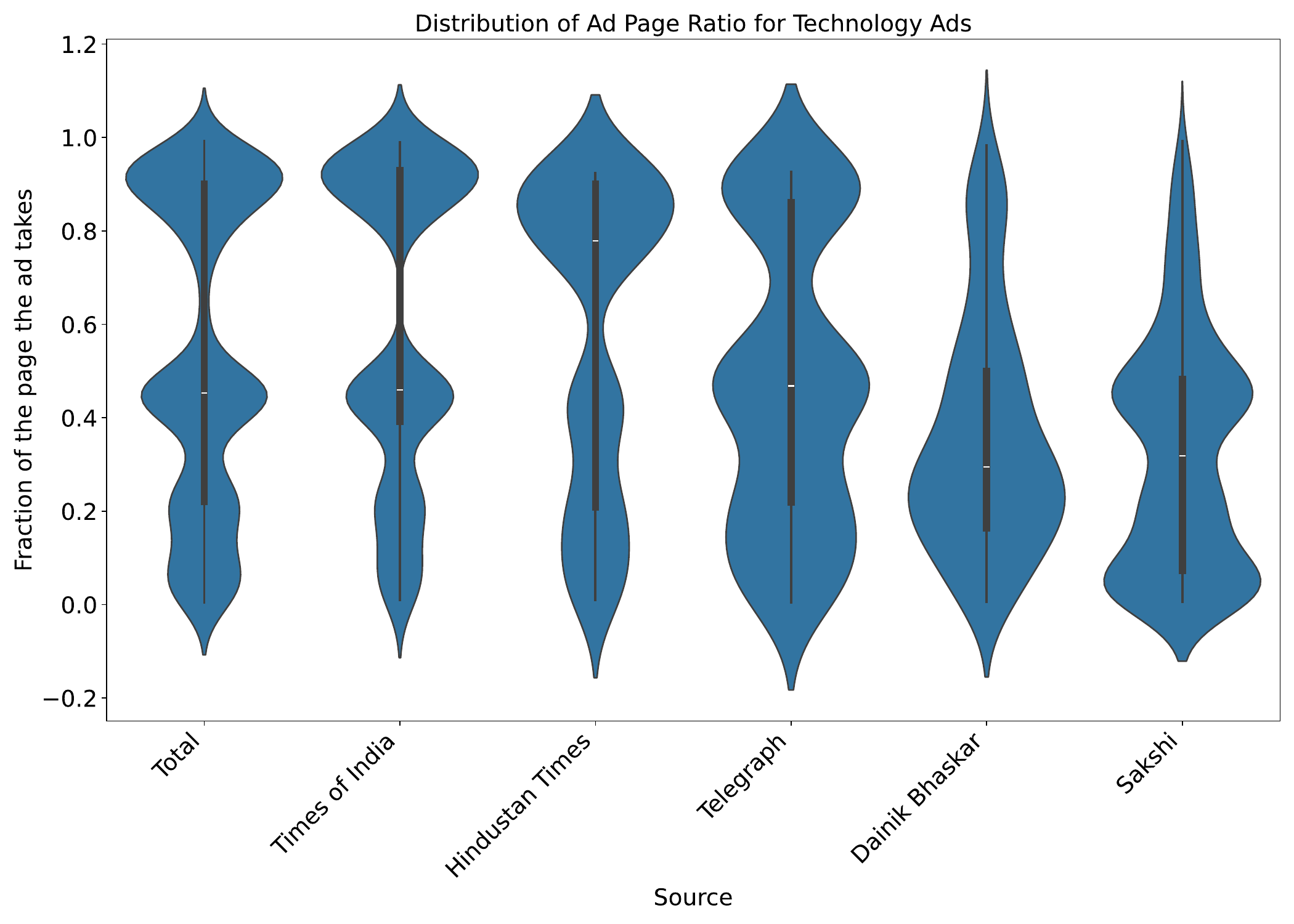}
    \caption{Technology ads across papers}
    \label{fig:technology_ads_across_papers}
\end{figure}

\begin{figure}[H]
    \centering
    \includegraphics[width=1\linewidth]{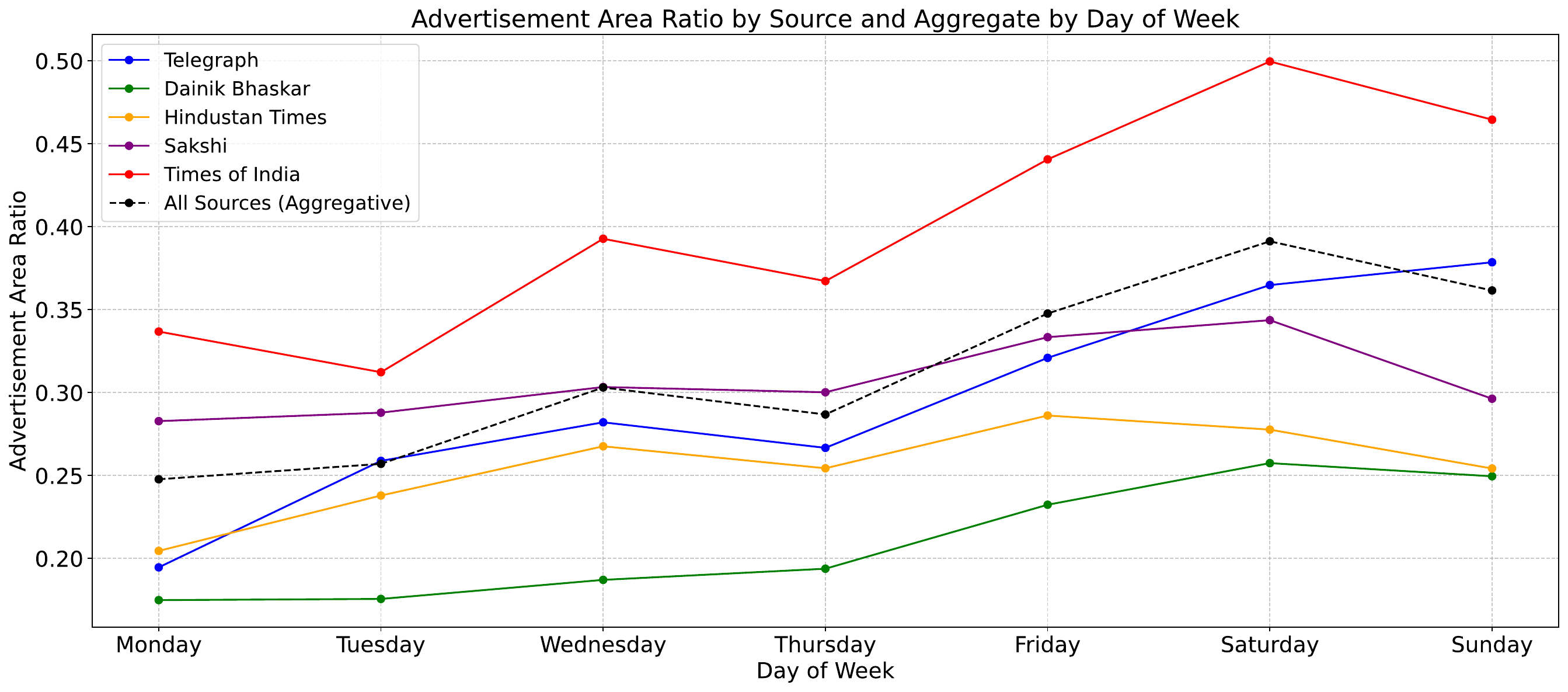}
    \caption{Advertiser Coverage by Day of the Week - Weekends show an increasing trend in Advertisement area}
    \label{fig:weekday}
\end{figure}

\begin{figure}[h]
    \centering
    \includegraphics[width=1\linewidth]{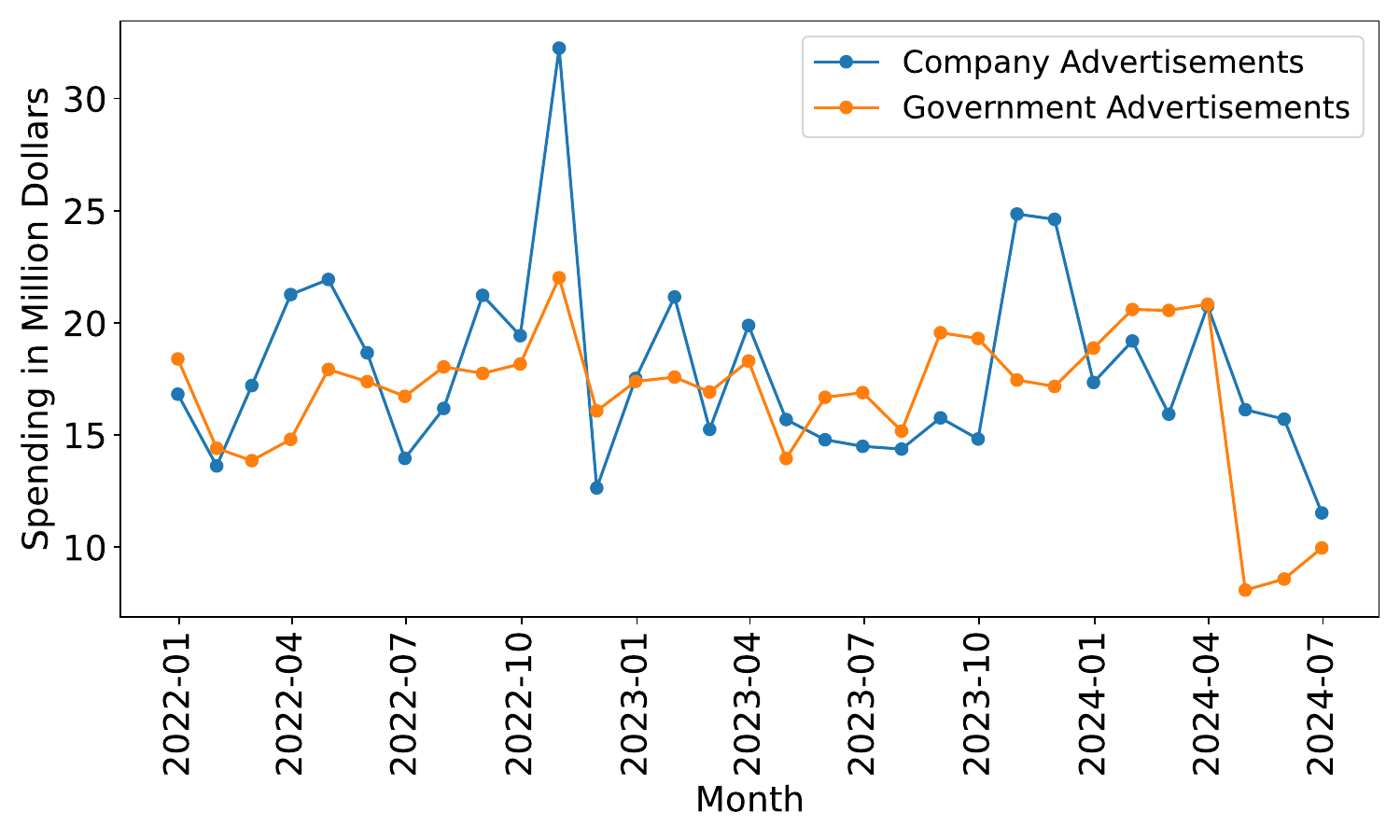}
    \caption{Spending by Companies and Government over Time.}
    \label{fig:ad_spend_temporal}
\end{figure}

In Figure \ref{fig:ad_spend_temporal}, we can see how both entities spend over time for a source, Times of India. Company spending exhibits clear seasonality, with fluctuations that may correspond to specific periods. Government spending appears more stable over time, with fewer pronounced peaks and stays close to company spending.

We can also observe the raw counts of articles and advertisements for various sources below.

\begin{figure}
    \centering
    \includegraphics[width=0.8\linewidth]{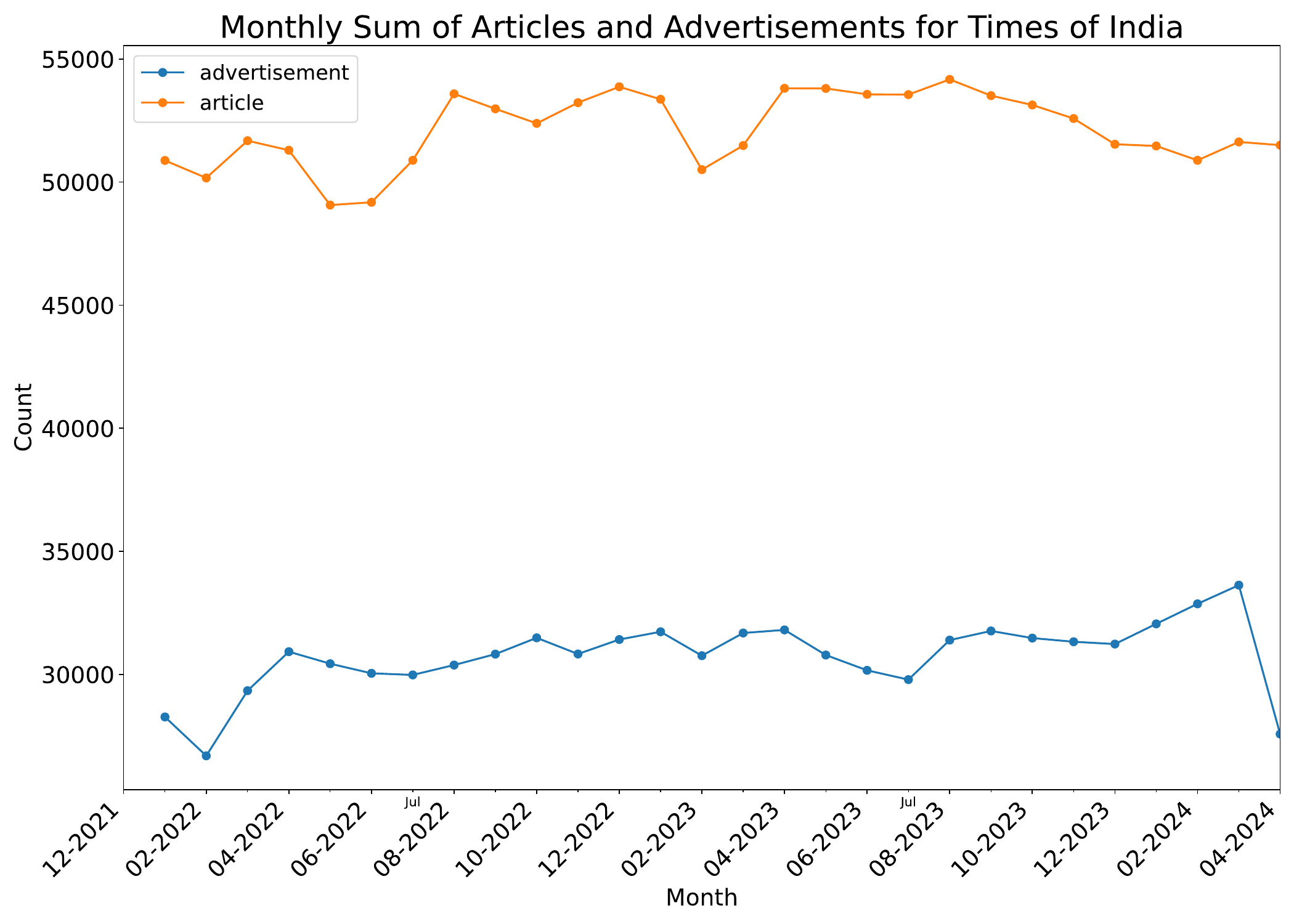}
    \caption{Count of Articles and Advertisements across time for Times of India}
    \label{fig:enter-label}
\end{figure}

\begin{figure}
    \centering
    \includegraphics[width=0.8\linewidth]{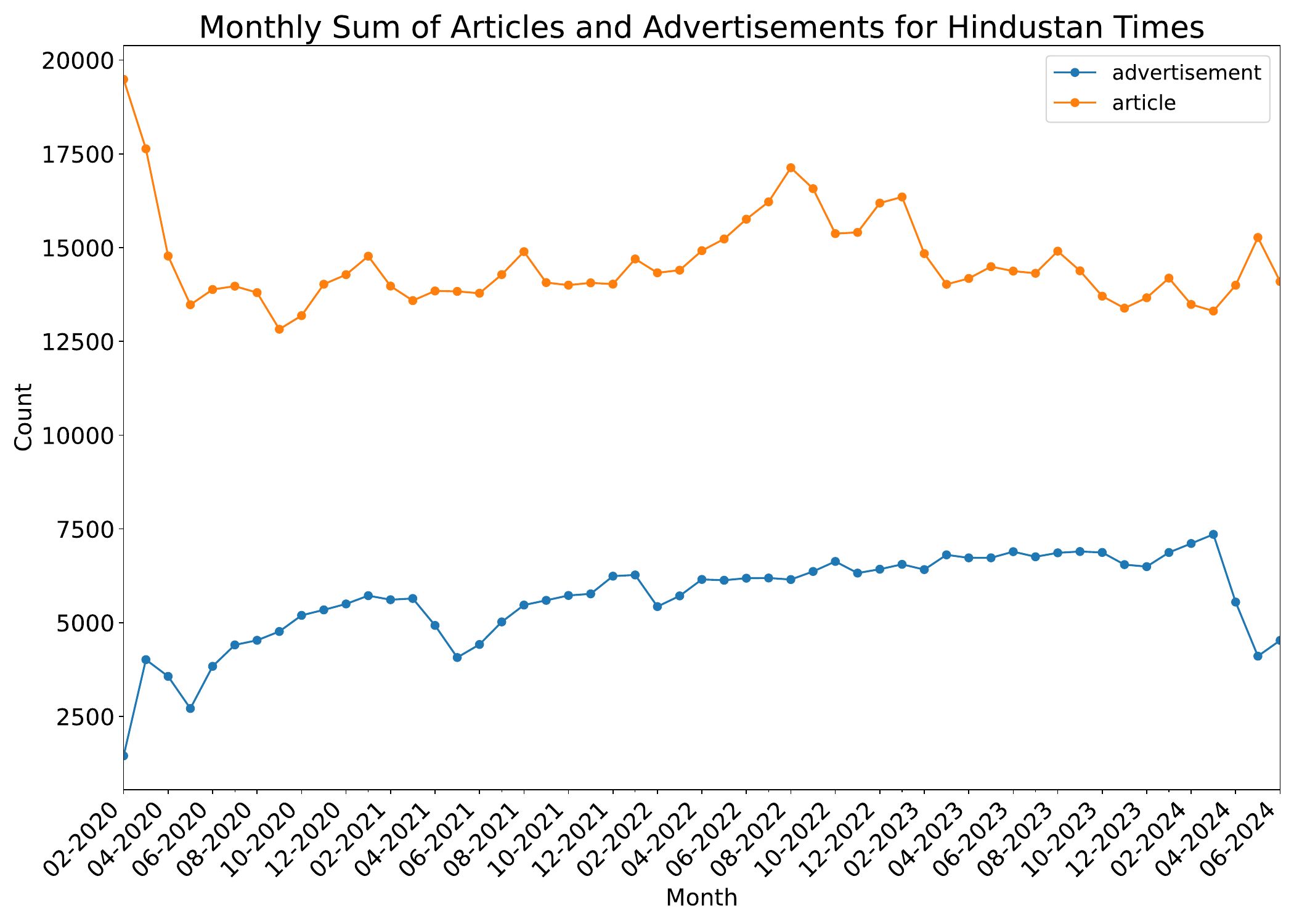}
    \caption{Count of Articles and Advertisements across time for Hindustan Times}
    \label{fig:enter-label}
\end{figure}

\begin{figure}
    \centering
    \includegraphics[width=0.8\linewidth]{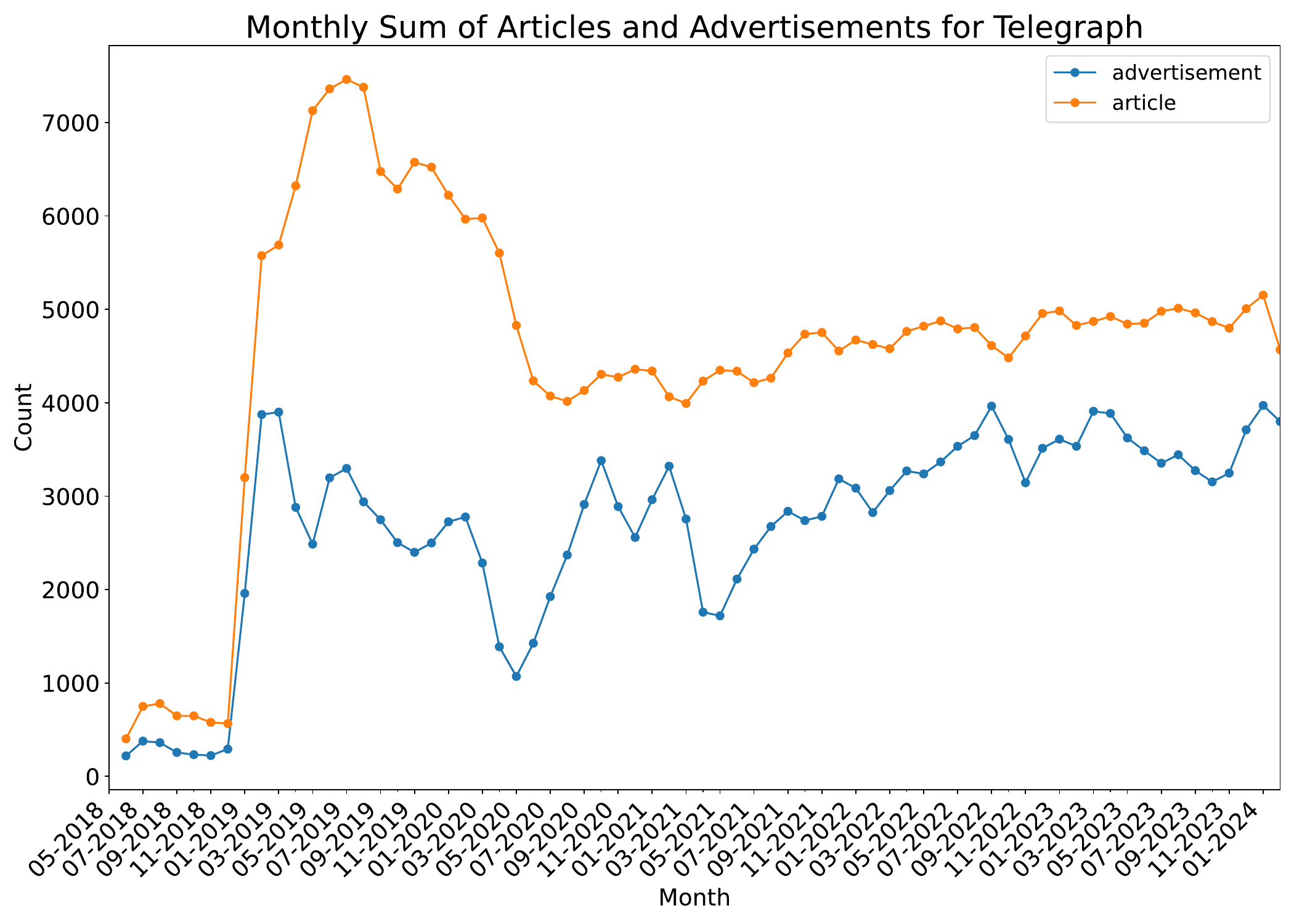}
    \caption{Count of Articles and Advertisements across time for Telegraph}
    \label{fig:enter-label}
\end{figure}

\begin{figure}
    \centering
    \includegraphics[width=0.8\linewidth]{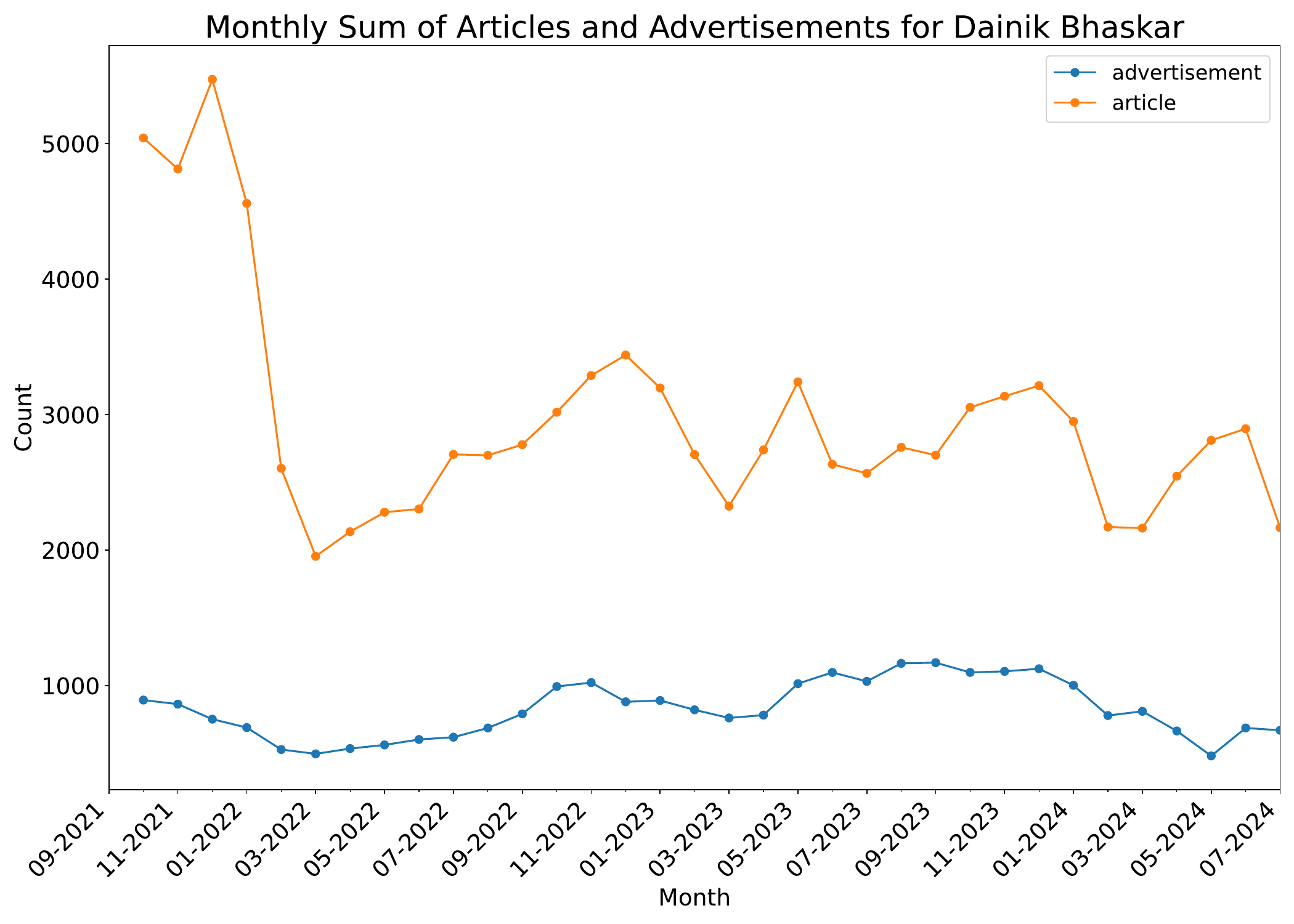}
    \caption{Count of Articles and Advertisements across time for Dainik Bhaskar}
    \label{fig:enter-label}
\end{figure}

\begin{figure}[H]
    \centering
    \includegraphics[width=0.8\linewidth]{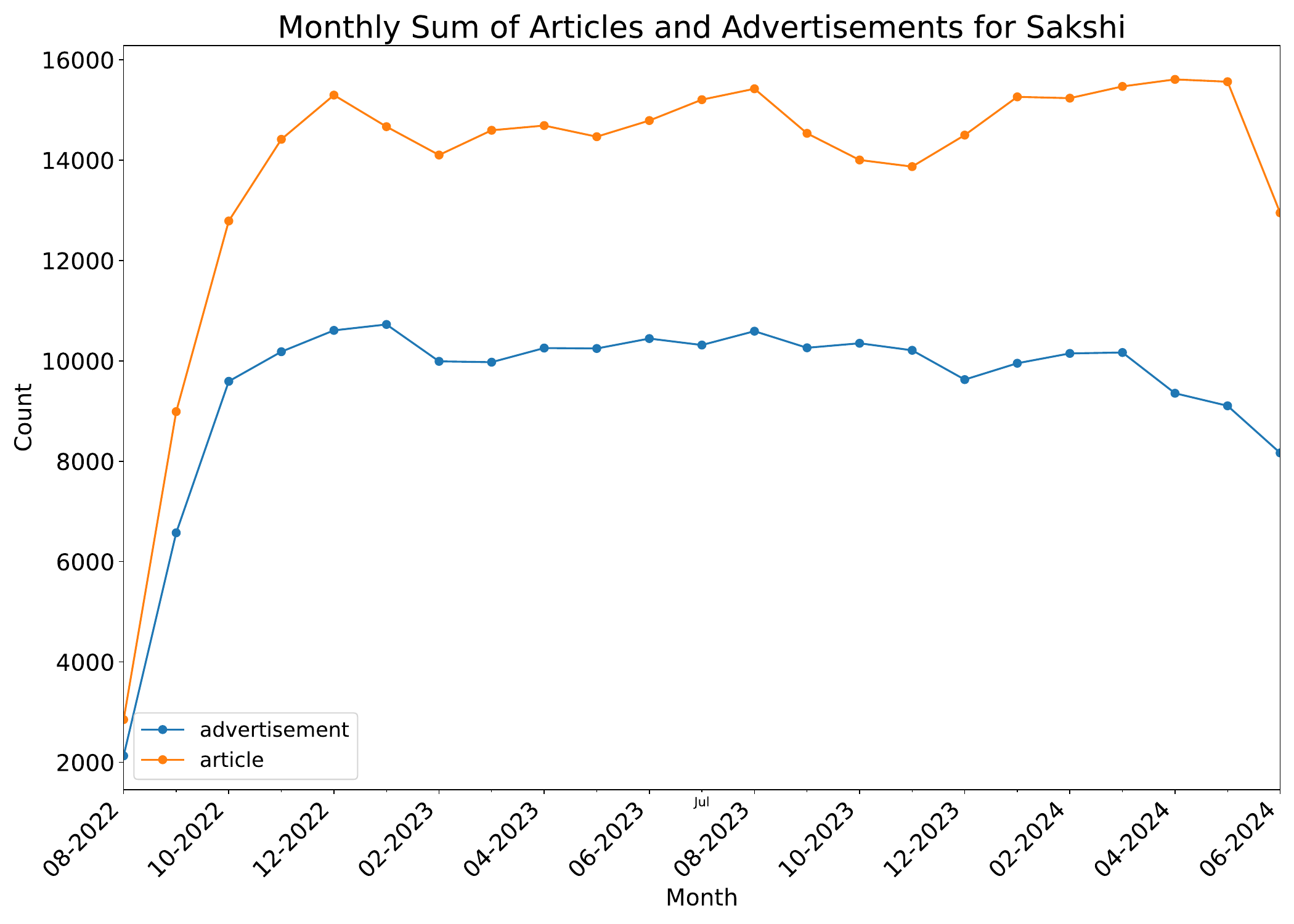}
    \caption{Count of Articles and Advertisements across time for Sakshi}
    \label{fig:enter-label}
\end{figure}

\section{Surya OCR in Reading Order}
\begin{figure*}
    \centering
    \includegraphics[width=0.7\linewidth]{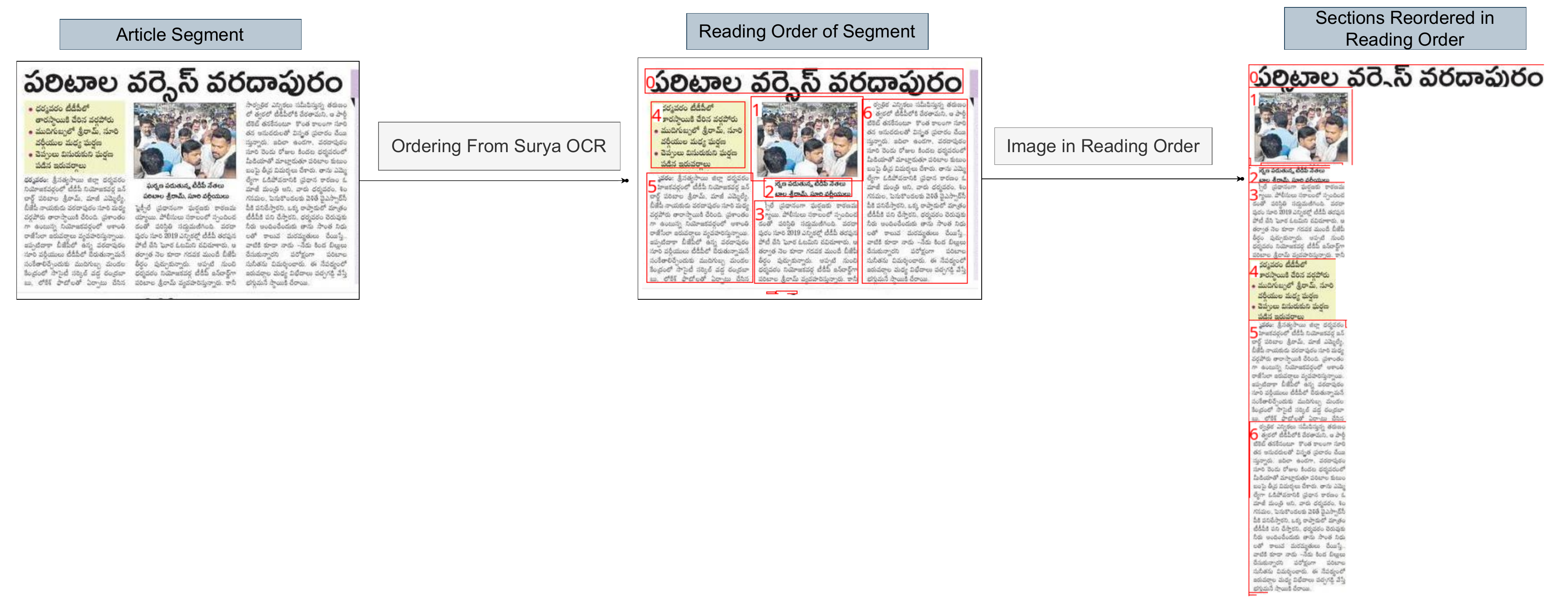}
    \caption{Processing Indic Image Segments}
    \label{fig:suryaocr}
\end{figure*}

For English Articles, We leverage Tesseract OCR and its Page Segmentation Modes to extract text in reading order, but due to no similar functionality in Surya OCR, We first generate a reading order from the image segment using Surya OCR, then process the image into a manner which is sequentially in reading order and then run Surya OCR on the processed image. A detailed run of the process can be seen in Fig ~\ref{fig:suryaocr}
\label{appendix_surya}

\section{Impact of Advertising and Scandals on Sentiment}
\begin{figure}[H]
    \centering
    \includegraphics[width=1\linewidth]{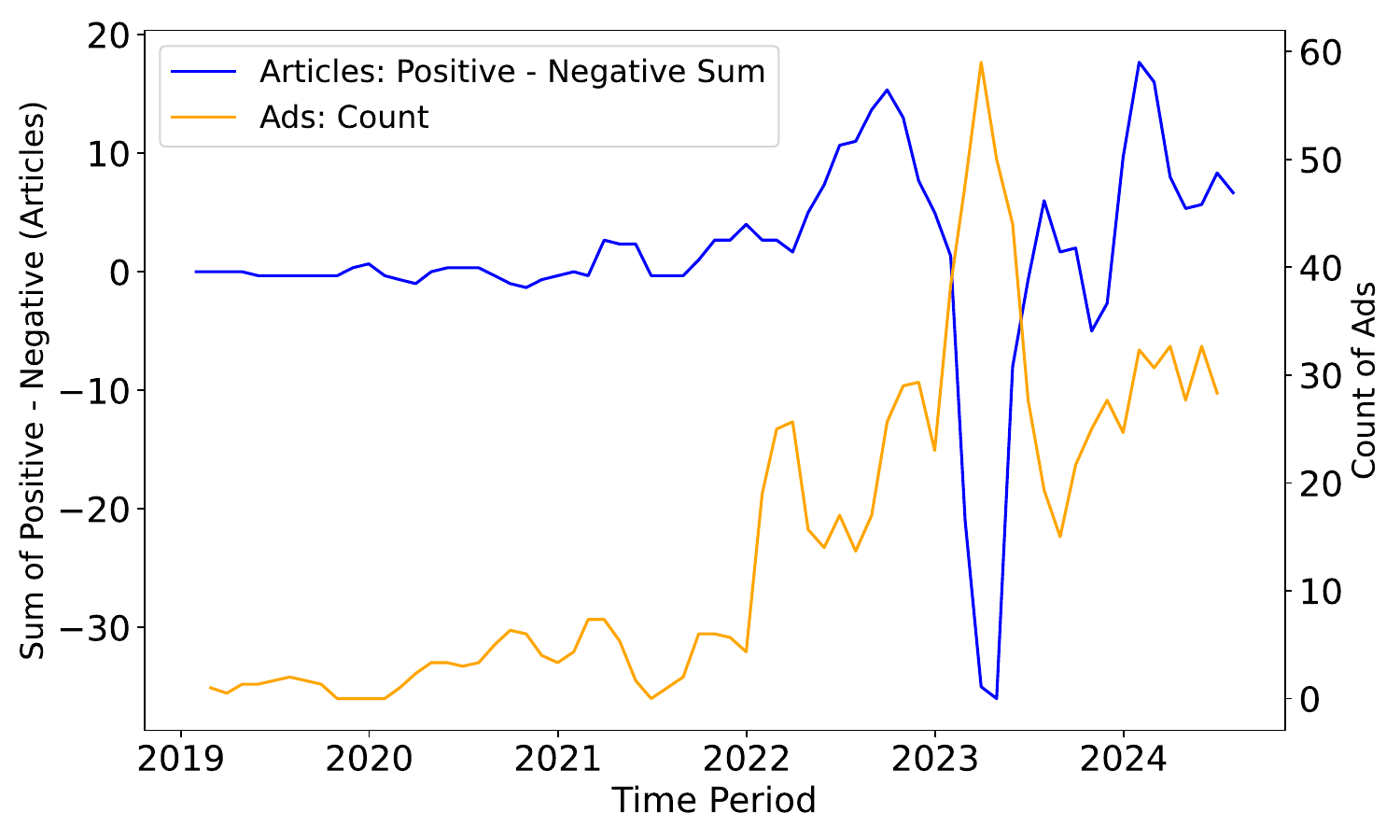}
    \caption{Influence of Advertising on Sentiment for Adani Conglomerate}
    \label{fig:adani}
\end{figure}

Figure \ref{fig:adani} clearly demonstrates the influence of advertising on sentiment, particularly in how scandals impact conglomerates. This conglomerate was significantly affected by negative media coverage following a scandal in Jan 2023, prompting an increase in advertising efforts to mitigate the fallout. The data highlights the response, which is buying more ads, and gradually, the sentiment shifts following a scandal and gradually returns to pre-scandal levels.

\section{Regression Tables}
The supplementary panel regressions on interactions with popularity with Corporate ads are provided in Tables ~\ref{tab:regression_popular_sentiment} and ~\ref{tab:regression_popular_counts}.

\begin{table*}[ht]
\centering
\caption{Panel Regression Results: The Impact of Ad Page Percentage and Popularity on Total Sentiment. Standard errors are clustered by entity.}
\label{tab:regression_popular_sentiment}
\begin{tabular}{lcccc}
\hline
 & (1) & (2) & (3) & (4) \\
\textbf{Dependent Variable: Total Sentiment} \\
\hline
\textbf{Total Ad Page Percent (Coefficient)} & -0.0003 & 0.0018 & -0.0016 & -0.0011 \\
 & (0.0031) & (0.0015) & (0.0019) & (0.0014) \\
\textbf{Popularity (Coefficient)} & 0.0032*** & 0.0014 & 0.0041** & 0.0004 \\
 & (0.0009) & (0.0017) & (0.0018) & (0.0013) \\
\textbf{Fixed Effects: Company} & No & Yes & No & Yes \\
\textbf{Fixed Effects: Time} & No & No & Yes & Yes \\
\hline
\textbf{Entity Count} & 40 & 40 & 40 & 40 \\
\textbf{Time Period Count} & 1372 & 1372 & 1372 & 1372 \\
\textbf{R\textsuperscript{2}} & 0.0625 & 0.0004 & 0.0110 & 7.399e-05 \\
\hline
\end{tabular}
\end{table*}

\begin{table*}[ht]
\centering
\caption{Panel Regression Results: The Impact of Ad Page Percentage and Popularity on Count Of Articles. Standard errors are clustered by entity.}
\label{tab:regression_popular_counts}
\begin{tabular}{lcccc}
\hline
 & (1) & (2) & (3) & (4) \\
\textbf{Dependent Variable: Count Of Articles} \\
\hline
\textbf{Total Ad Page Percent (Coefficient)} & 0.0640*** & 0.0185*** & 0.0299*** & 0.0214*** \\
 & (0.0154) & (0.0042) & (0.0087) & (0.0078) \\
\textbf{Popularity (Coefficient)} & 0.0246*** & 0.0109 & -0.0163*** & 0.0136 \\
 & (0.0041) & (0.0069) & (0.0061) & (0.0079) \\
\textbf{Fixed Effects: Company} & No & Yes & No & Yes \\
\textbf{Fixed Effects: Time} & No & No & Yes & Yes \\
\hline
\textbf{Entity Count} & 40 & 40 & 40 & 40 \\
\textbf{Time Period Count} & 1372 & 1372 & 1372 & 1372 \\
\textbf{R\textsuperscript{2}} & 0.5649 & 0.0118 & 0.0686 & 0.0150 \\
\hline
\end{tabular}
\end{table*}

\section{Price Variations and Scaling Factor}
\label{appendix_price_page}
To better account for price variations across different pages and provide a more accurate measure of advertising expenditures, we apply a scaling factor to pages of particular interest, where advertising costs are typically higher. These pages include the first page, the third page, and the last page, all of which command premium rates. Additionally, we gather city-specific and source-specific advertising rates from online sources. To facilitate regression analysis and make the coefficients more interpretable, each entry is normalized by dividing by the minimum rate (546) in the dataset. This normalization allows for a clearer examination of the relationship between advertising expenditures and media coverage. The rates used for the table can be found here for the \textbf{Times of India}\footnote{\url{https://web.archive.org/web/20241001131208/https://riyoadvertising.com/times-of-india-display-ad-rates.html}}, \textbf{Hindustan Times}\footnote{\url{https://web.archive.org/web/20241001131804/https://www.hindustantimes.com/rate-card/Impactht}}, \textbf{The Telegraph}\footnote{\url{https://www.bhavesads.com/the-telegraph/display-ad-rates}}, \textbf{Dainik Bhaskar}\footnote{\url{https://web.archive.org/web/20240503230534/http://www.riyoadvertising.com/dainik-bhaskar.html}}, and \textbf{Sakshi}\footnote{\url{https://web.archive.org/web/20241001134917/https://riyoadvertising.com/sakshi.html}}.

\begin{table*}[h]
    \centering
\caption{Page rates for various papers.}
\label{tab:page_rates}
    \begin{tabular}{l|l|l|l|l|l}
    \hline
        \textbf{Source} & \textbf{City} & \textbf{1st Page} & \textbf{3rd Page} & \textbf{Back Page} & \textbf{Base Price} \\ \hline
        Times of India & Mumbai & 9665 & 6850 & 7230 & 5640 \\ \hline
        Times of India & Delhi & 6355 & 4830 & 5075 & 4120 \\ \hline
        Times of India & Kolkata & 2435 & 1920 & 2105 & 1835 \\ \hline
        Times of India & Chennai & 2815 & 2381 & 2405 & 1985 \\ \hline
        Hindustan Times & Mumbai & 5100 & 3750 & 3750 & 3000 \\ \hline
        Hindustan Times & Delhi & 10750 & 7470 & 7470 & 5970 \\ \hline
        Dainik Bhaskar & Delhi & 867 & 661 & 774 & 546 \\ \hline
        Sakshi & Andhra & 6739 & 2995 & 5990 & 2995 \\ \hline
        Sakshi & Hyderabad & 2700 & 1200 & 2400 & 1200 \\ \hline
        Sakshi & Telangana & 2700 & 1200 & 2400 & 1200 \\ \hline
        Telegraph & Kolkata & 2641 & 2565 & 2430 & 2230 \\ \hline
    \end{tabular}
\end{table*}

\section{Keywords To Identify Government-based articles and advertisements}
The keywords used to identify Government-based articles and ads can be found in Table \ref{tab:ad_corruption_keywords}. 
Articles are identified if there is a match with any keyword from the corruption-related list and the presence of a government-related term ("government," "govt," "state," or "central") to ensure government corruption-related articles. In contrast, advertisements are classified if they contain any keyword from the advertisement-related list. 
\label{govt_keywords}
\begin{table*}
    \centering
    \caption{Advertisement and Corruption Keywords.}
    \label{tab:ad_corruption_keywords}
    
    \begin{tabular}{p{5cm}|p{10cm}}
    \hline
        \textbf{Category} & \textbf{Keywords} \\ \hline
        Advertisement Keywords & government, state, central, tender, gov, e-tender, corrigendum, e-corrigendum, govt., tenders, procurement, e-procurement \\ \hline
        Corruption Keywords & bail, black, bribe, cbi, chor, conspiracy, corrupt, croni, demonet, expos, helicopt, investig, jail, jumla, lokpal, loot, nirav, probe, prosecut, rafal, raid, scam, scandal, steal, theft, thief, illegal, fraud, embezzle, misappropriat, laundering, offshore, tax evasion \\ \hline
    \end{tabular}
\end{table*}


\section{Keywords To Identify Companies}
\label{company_keywords}
The keywords are mostly curated by adding subsidiaries and brands that are uniquely identifiable to a brand. 
Table~\ref{tab:company_keywords}.

\begin{table*}[!ht]
    \centering
\caption{Company keywords.}
\label{tab:company_keywords}

    \begin{tabular}{p{3cm}|p{12cm}} 
    \hline
        \textbf{Company} & \textbf{Keywords} \\ \hline
        Tata &  Tata , Jaguar Land Rover , Taj Hotels , BigBasket , 1mg , AirAsia , Vistara , Tanishq , Titan , Starbucks , Voltas , Vivanta , Air India , Croma  \\ \hline
        Reliance &  Reliance , JioFiber , JioMart , AJIO , Netmeds , Hamleys , Urban Ladder  \\ \hline
        Hindustan Unilever &  Hindustan Unilever , HUL , Lakmé , Lifebuoy , Dove , Surf Excel , Kwality Wall's , Bru , Kissan , Vaseline , Ponds , Pepsodent , Clinic Plus , Rin , Axe  \\ \hline
        Procter \& Gamble (P\&G) &  Procter \& Gamble , Procter and Gamble , P\&G , Pampers , Ariel , Tide , Gillette , Whisper , Vicks , Olay , Pantene , Head \& Shoulders , Oral-B , Old Spice  \\ \hline
        ITC Limited &  ITC Limited , Sunfeast , Aashirvaad , Savlon , Fiama , Vivel , ITC Hotels , Bingo! , Yippee! , Classmate , Wills , Gold Flake  \\ \hline
        Godrej Group &  Godrej , Good Knight , Cinthol  \\ \hline
        Bharti Airtel &  Airtel , Wynk Music  \\ \hline
        Samsung &  Samsung  \\ \hline
        Xiaomi &  Xiaomi , Redmi , POCO , Mi TV , Mi Smart Home , Mi Ecosystem , MIUI  \\ \hline
        Vivo &  Vivo  \\ \hline
        Oppo &  Realme , Oppo  \\ \hline
        OnePlus &  OnePlus  \\ \hline
        Maruti Suzuki &  Maruti Suzuki , Suzuki , Nexa  \\ \hline
        LIC &  Life Insurance Corporation of India , LIC  \\ \hline
        Hyundai &  Hyundai Motor India , Kia , Hyundai  \\ \hline
        Toyota Kirloskar &  Toyota  \\ \hline
        Renault India &  Renault , Dacia  \\ \hline
        MG Motor India &  Morris Garages , MG Hector , MG Motor  \\ \hline
        Stellantis &  Stellantis , Jeep India , Citroen India , Fiat , Mopar  \\ \hline
        BMW Group India &  BMW  \\ \hline
        Mercedes-Benz India &  Mercedes-Benz  \\ \hline
        Amazon &  Amazon , Kindle  \\ \hline
        Coca-Cola &  Thums Up , Sprite , Fanta , Minute Maid , Kinley , Maaza , Coca-Cola , Diet Coke , Smartwater  \\ \hline
        PepsiCo &  PepsiCo , Pepsi , Mirinda , 7Up , Lay's , KurKure , Tropicana , Mountain Dew , Gatorade , Quaker Oats  \\ \hline
        Adani Group &  Adani , Ambuja Cements  \\ \hline
        Mahindra Group &  Mahindra \& Mahindra , Mahindra Tractors , Tech Mahindra , Mahindra Finance , Mahindra Electric , Club Mahindra , Mahindra Lifespaces , Automobili Pininfarina  \\ \hline
        Nestlé &  Nestlé India , Maggi , Nescafé , KitKat , Milo , Milkmaid , Nestea , Cerelac , Everyday , Perrier  \\ \hline
        Sony &  Sony , PlayStation , SonyLIV  \\ \hline
        Volkswagen Group &  Volkswagen , Audi , Porsche , Bentley , Lamborghini , Skoda , Bugatti , Ducati  \\ \hline
        Ford Motor Company &  Ford  \\ \hline
        Apple &  iPhone , iPad , MacBook , Apple Watch , iMac , Apple TV , Apple Music , Apple Pay , iCloud , Apple Store  \\ \hline
        Google (Alphabet) &  Google Search , YouTube , Google Maps , Google Cloud , Google Ads , Android , Google Play , Gmail , Google Pixel , Nest  \\ \hline
        Hero MotoCorp &  Hero MotoCorp , Splendor , HF Deluxe , Passion , Glamour , Xpulse , Hero MotoSports  \\ \hline
        Honda Motorcycle and Scooter India &  Honda , Activa  \\ \hline
        Bajaj Auto &  Bajaj , Pulsar , Dominar , Avenger , KTM , Husqvarna , Bajaj Finserv , Bajaj Finance  \\ \hline
        FIITJEE &  FIITJEE  \\ \hline
        Byju's Aakash &  Byju's , Aakash  \\ \hline
        Allen Career Institute &  Allen  \\ \hline
        Nissan &  Nissan , Datsun  \\ \hline
        Prestige &  Prestige TTK  \\ \hline
        BigBasket &  BigBasket , BB Daily , BBinstant  \\ \hline
        Amul &  Amul  \\ \hline
        Patanjali &  Patanjali  \\ \hline
    \end{tabular}
\end{table*}